\documentclass{article}
\font\cero=cmss10 scaled 1728 
\usepackage{tensor}
\usepackage{verbatim}
\usepackage{amsfonts}
\usepackage{graphicx}
\usepackage{amssymb}
\usepackage{float}
\begin{document}
\begin{flushleft}
{\cero Spontaneous symmetry breaking, and strings defects in hypercomplex gauge field theories}\\
\end{flushleft}
{\sf R. Cartas-Fuentevilla}\\
{\it Instituto de F\'{\i}sica, Universidad Aut\'onoma de Puebla,
Apartado postal J-48 72570 Puebla Pue., M\'exico}; 

E-mail: rcartas@ifuap.buap.mx

\noindent{\sf O. Meza-Aldama} \\
{\it Facultad de Ciencias F\'{\i}sico-Matem\'{a}ticas, Universidad Aut\'onoma de Puebla,
Apartado postal 1152, 72001  Puebla Pue., M\'exico.} \\

ABSTRACT: Inspired by the appearance of split-complex structures in the dimensional reduction of string theory, and in the theories emerging as byproducts, we study the hyper-complex formulation of Abelian gauge field theories, by incorporating a new complex unit to the usual complex one.  The hypercomplex version of the traditional Mexican hat potential associated with the $U(1)$ gauge field theory, corresponds to a {\it hybrid} potential with two real components, and with $U(1)\times SO(1,1)$ as symmetry group. Each component corresponds to a 
deformation of the hat potential, with the appearance of a new degenerate vacuum. Hypercomplex electrodynamics will show novel 
properties, such as the spontaneous symmetry breaking scenarios with running masses for the vectorial and scalar Higgs fields,  and the Aharonov-Bohm type strings defects as exact solutions; these topological defects may be detected only by quantum interference of charged particles through gauge invariant loop integrals.  In a particular limit, the {\it hyperbolic} electrodynamics does not admit topological defects associated with continuous symmetries.\\


\section {Introduction}
\label{intro}
Explorations involving hypercomplex structures have appeared recently in the literature, for example, in the dimensional reduction of M-theory over a Calabi-Yau-3 fold, where a five-dimensional ${\cal N}=2$ supergravity theory emerges, it turns out that the hyperbolic representation based on para- or split-complex  numbers is the most natural way to formulate the scalar fields of the five-dimensional universal multiplet, gaining insight in the understanding of the string theory landscape \cite{emam,emam1}. In this context, the switching on of the split-complex form of the theory, solves automatically the inconsistencies related with the finding of well-behaved solutions representing the so called BPS instantons and 3-branes. In the same context,  the Lagrangian and the supersymmetric rules used in \cite{perry} for a description of the so called D-instantons in terms of supergravity, require by consistency of a substitution rule that changes  the standard imaginary unit $i$ with $i^2=-1$,  by a formal new imaginary  unit $j$ with $j^2=1$, which is different from $\pm 1$, and corresponds to an algebra of para-complex numbers. The formal description of such a mysterious substitution rule is given in \cite{cortes,cortes1} in terms of para-complex manifolds endowed with a special para-K\"ahler geometry, with applications in the study of instantons, solitons, and cosmological solutions in supergravity and M-theory. 

In a different context, the para-complex numbers appear as a hyperbolic unitary extension of the usual complex phase symmetry of electromagnetism in order to generalize it to a gravito-electromagnetic gauge symmetry \cite{U}. Additionally an alternative representation of relativistic field theories is given in \cite{U1,U2,JH} in terms of hyperbolic numbers; in particular the Dirac equation and the Maxwell equations admit naturally such a representation, in which one has both the ordinary and the hyperbolic imaginary units; along the same lines it is shown that the $(1+1)$ string world-sheet possesses an inherent hyperbolic complex structure \cite{JH}.
Furthermore, in \cite{U3} the requirement of hermiticity on the Poincar\'e mass operator defined on the commutative ring of the hyperbolic numbers ${\cal H}$, leads to a decomposition of the corresponding hyperbolic Hilbert space into a direct product of the Lorentz group related to the spacetime symmetries, and the hyperbolic unitary group  $SU(4,H)$, which is considered as an internal symmetry of the relativistic quantum state; the hyperbolic unitary group is equivalent to the group $SU(4,C) \times SU(4,C)$ of the Pati-Salam model \cite{PS}.  In \cite{U4} the hyperbolic Klein-Gordon equation for fermions and bosons is considered as a para-complex extension of groups and algebras formulated in terms of the product of ordinary complex and hyperbolic unit;  this implies the existence of hyperbolic complex gauge transformations, and the possibility of new interactions; however, although certainly there is not currently experimental indications of them, either evidence against. If these new interactions are effectively absence, then it is of interest to understand why the hyperbolic complex counterparts for the other interactions there no exist in nature at presently known energies, in spite of the consistence of hyperbolic extensions from the theoretical point of view. However, also it raises the interesting possibility of realizations of those counterparts beyond presently known energies, including those close to the Planck scale.

On the other hand, the presence of hyperbolic phases implies symmetries associated with non-compact gauge groups; these symmetries can be realized as symmetries of the background spacetime and/or internal gauge symmetries; as example, it is well known the appearance of non compact internal symmetries in the context of gravity, with the invariance under diffeomorphisms as the fundamental symmetry gauge. Furthermore, although gauge theories are typically discussed for compact gauge groups, the  integrable sectors of QDC,  ghost-, and $\theta$-sectors manifest the presence of non-compact gauge groups  \cite{noc}, with surprising new features.
For example, the spontaneous symmetry breaking is possible in low dimensions provided that non-compact groups are present, evading the Mermin-Wagner theorem (\cite{seiler}, and references therein); at quantum level, the Hilbert space is nonseparable. Similarly in the study of quantum non-compact $\sigma$ models, as opposed to the case of quantum electrodynamics, the theory can be correctly quantized only in a Hilbert space with indefinite metric \cite{nojiri}; in the case of a positive-definite Hilbert space, the quantization requires an extended space  that incorporates negative-energy modes \cite{holten}. In general the classical and quantum descriptions of noncompact $\sigma$ models show problems such as the unitarity of the $S$ matrix, and the spontaneous symmetry breaking realizations. 

\section{ Motivations and an advance of results}
\label{moti}
One of the motivations of the present work is to explore the realizations of the hyperbolic symmetries as an internal gauge symmetry in classical gauge field theories; we determine the effects of the incorporation of those symmetries on the geometry and topology of the vacuum manifolds, and the subsequent effect on the formation of topological defects. We shall find that the switching on the hyperbolic structures, gives the possibility that the degenerate vacuum in gauge theories can correspond to non-compact manifolds;
the non-compact character of the vacuum will have a nontrivial effect on the possible formation of topological defects associated to continuous symmetries through the Kibble mechanism, when the new hyperbolic symmetry breaks down. Conveniently interpreted, these results will be consistent with the lack of strong and convincing evidences of the existence of cosmic topological defects, and consequently with the possibility that the hyperbolic symmetry is present at some moment in the sequence of symmetry breakdown in the early-universe phase transitions. Incidentally the incorporation of non-compact gauge groups will allow us to gain insight in certain aspects of gravity theories from the perspective of a deformed version of conventional gauge theories.

 Topological structures have been object of intense research  due to their relevance in confinement and quiral symmetry breaking phenomena
in quantum chromodynamics; it is well known the role that the monopoles may play as a possible source of confinement \cite{mandel, tft};
more recently the so called dyons have been considered as alternative source for such a phenomenon \cite{facci}. In general the topological structure of any gauge theory is conditioned by the existence of non-trivial homotopy groups, and these are determined by the specific symmetry gauge groups and their stability subgroups; recently the Weyl symmetric structure of the classical QCD vacuum is described by a second homotopy group constraint, which determines the monopole charge \cite{cho}. Similarly, the knot topology of QCD vacuum is determined by a third homotopy group constraint \cite{knot};  this approach is useful in the construction of new analytic solutions. With these motivations, in this work we determine the effects of the incorporation of hyperbolic rotations as a part of the internal gauge group that usually involves only compact gauge groups, on the topological structures of the theories considered.

In the Section \ref{hyrot} we consider the hyper-complex numbers with the purpose of introducing the hyperbolic phases as a part of the symmetry gauge group, that in general will include the usual $U(1)$ compact phases. Then, the hypercomplex deformation with global phases of the classical massive $\lambda\phi^4$ model is developed in Section \ref{lambda4}; in particular in the Section \ref{gamma1} the purely hyperbolic version of the traditional mexican hat potential is analyzed; hyperbolic version means the substitution of the usual complex unit $i$ by $j$. This theory does not allow topological defects associated with continuous symmetry. In the Sections  \ref{gammano1}, and  \ref{nogamma11} the hyperbolic deformations of the Mexican hat potential is developed; deformation will imply the incorporation of the new complex unit $j$, to the usual unit $i$. 
In Section \ref{vm} we describe geometrical and topologically the vacuum manifold, which will correspond to a two-dimensional non-compact space embedded in the four dimensional hypercomplex space as ambient space; the homotopy constraints are analyzed.
The polar description of the spontaneous symmetry breakdown is developed in Section \ref{polar};
the polar parametrization for the fields will allow to describe circular, hyperbolic, and radial oscillations, and we shall make a comparison with the usual treatments that involve only compact gauge groups. 
The formation of possible topological defects are analyzed in Section \ref{topdef}; in this section the Derrick's theorem is confirmed.
Furthermore, in the case of theories with $U(1)$ gauge symmetries, the conventional vacuum manifold is defined by the bottom of the {\it mexican hat} potential, the circle; in the formulation at hand that circle will be retained as a compact transversal section of the new non-compact vacuum manifold that incorporates the hyperbolic phases.

Finally in Section \ref{hed} the circular and hyperbolic local rotations are considered by the coupling to hypercomplex electrodynamics;  this theory leads to spontaneous symmetry breaking scenarios with hypercomplex scalar and vectorial fields with running masses which mimic the flows of the renormalization groups. In particular, in Section \ref{polar1} a purely massive electrodynamics, without the presence of scalar Higgs fields is obtained. In Section \ref{tdlocal} the local topological strings are studied; these defects will turn out to be of the Aharonov-Bohm type, detectable only by quantum interference of charged particles, in consistency with previous studies on the subject; this issue is discussed in detail in Section \ref{bohm} in concluding remarks. We speculate at the end, on the possible cosmological implications.

\section {Incorporating the hyperbolic rotations}
\label{hyrot}
As an extension of the conventional complex numbers, the commutative ring of hypercomplex numbers, $z \in {\cal H}$ are defined as \cite{U},
\begin{eqnarray}
     z =  x + iy +jv + ijw, \quad \overline{z}  =  x - iy -jv + ijw, \quad  x,y,v,w \in {\cal R} 
          \label{ring}
\end{eqnarray} where the hyperbolic unit $j$ has the properties $j^{2}=1$, and $\overline{j} = -j$, and, as usual, $i^{2}=-1$, and $\overline{i}=-i$. Hence, with respect to the conjugation involving both complex units, the square of the hypercomplex number is given by
\begin{equation}
     z \overline{z} = x^{2} + y^{2} -v^{2} - w^{2} +2ij (xw-yv),
     \label{square}
\end{equation} 
which is not a real number, instead it is in general a {\it Hermitian} number. 
The expression (\ref{square}) is invariant under the usual circular rotations $e^{i\theta}$ represented by the Lie group $U(1)$; similarly it is invariant under {\it hyperbolic} rotations that can be represented by the connected component of the Lie group $SO(1,1)$ containing the group unit. A hyperbolic rotation is represented by the hyperbolic versor $e^{j\chi} \equiv \cosh\chi + j \sinh\chi$, with the split-complex conjugate $e^{-j\chi} = \cosh\chi - j\sinh\chi$, and with the operations $e^{j\chi} \cdot e^{j\chi'} = e^{j(\chi +\chi')}$.The hyperbolic rotations correspond to a subgroup of the group $SL(2,R)$, which represents all linear transformations of the plane that preserve oriented area. The elements of the group are classified as elliptic, parabolic, or hyperbolic, depending on whether the $|trace|<2$, $|trace|=2$, or $|trace|>2$ respectively. The respective subgroups are obtained by incorporating $\pm I$, with $I$ the identity element; in particular, the elements of the hyperbolic subgroup are identified with {\it squeeze} mappings, which correspond geometrically to preserving hyperbolae in the plane, with the hyperbolic angle playing the role of invariant measure of the subgroup.
Since the image points of the squeeze mapping are on the same hyperbola, then such a mapping preserves the form $x \cdot y$, and can be identified with a {\it hyperbolic rotation} in analogy with circular rotations preserving circles. The hyperbolic subgroup will play a central role in this paper, since the invariance under its action will be revealed as a fundamental internal symmetry in field theory, due to that the invariant form $x \cdot y$ appears recurrently in physics. 

If $h$ is a hyperbolic element, then $|trace(h)|>2$, and $det(h)=1$, and can be parametrized as
\begin{equation}
     h = \left( \begin{array}{cc}
                 \eta e^{j\chi} & 0 \\
                 0 & \eta e^{-j\chi}
                 \end{array} \right), \qquad \eta =\pm 1, \qquad \chi\in R-\{ 0\} ;
     \label{hh}
\end{equation}
the identity element can be incorporated by allowing that $\chi =0$, and with the choice $\eta =1$; similarly with $\eta =-1$ the element $-I$ is added. Hence, the expression (\ref{hh}) parametrizes the elements of the hyperbolic subgroup  with $\chi\in R$; the part connected to the identity will be denoted by $SO^{+}(1,1)$, and represents the subgroup of continuous transformations; the discrete transformation related to the element $-I$ will act separately as a ${\cal PT}$-like transformation. Note that the hyperbolic subgroup is an Abelian group, and $SO^{+}(1,1)$ corresponds to an one-parameter Lie group, with a non-compact generator. Therefore, the quadratic form   (\ref{square}) is invariant under the full phase $e^{i\theta}e^{j\chi}$, corresponding to the group $U(1)\times SO(1,1)$. This symmetry largely ignored in the literature, will have a direct impact in various directions, in particular in the vacuum structure in theories with gauge symmetry, leading to radical changes in its topology and geometry.

The full non-compact group $SL(2,R)$ already has been considered previously as a structure group in a toy model; the kinematics of
the $SL(2,R)$ ``Yang-Mills" theory in 1+1 dimensions was studied in \cite{gote}; such an analysis was motivated in part for gaining insight in the formulation of gauge theories with non-compact structural groups, which may shed light in the quantization of any theory of gravity. The analysis shown that the configuration space has a non-Hausdorff ``network" topology, rather that a conventional manifold, and the emergent quantization ambiguity can not be resolved as opposed to the usual compact case. This toy model captures the relevant aspects of a four-dimensional non-compact Yang-Mills theory, which is physically more close to four dimensional gravity. In that analysis the foliation of the group $SL(2,R)$ by its conjugacy classes is used; the space of conjugacy classes associated with the elliptic and hyperbolic subgroups has the non-compact topology of a two-sheet hyperboloid
. A two sheet hyperboloid has a circle and a hyperbola as factor spaces, with trivial homotopy groups, except $\pi_{0}=2$; although in the present work  the vacuum manifold will have the circle and the hyperbola as factor spaces, the homotopy groups will be nontrivial.

However, for our purposes, we need to restrict ourselves to hypercomplex numbers that poseess only two degrees of freedom, {\it i.e.}, defined only in terms of two real quantities, in order to obtain a minimal deformation of an ordinary complex number that encodes two real quantities. Such a deformed version will incorporate the hyperbolic rotations to the usual $U(1)$-circular rotations in field theory; this requires the identification of the four real variables in Eq.\ (\ref{ring}) to each other.  The first case corresponds to 
the restrictions $x=\alpha v$, and $y=\alpha w$, with the parameter $\alpha\in {\cal R}$, which lead to the hypercomplex number
\begin{equation}
     z = (\alpha-j)(v-iw), \qquad     z\overline{z} = (\alpha^{2}-1) (v^{2} + w^{2}),
     \label{pd}
\end{equation} 
which has a square positive (negative) definite for  $\alpha^{2}>1$ ($\alpha^2<1$) ; note that a number of the form (\ref{pd}) is zero if and only if $v=0=w$.

Furthermore, a number of the form (\ref{pd}), $z_{1}=(\alpha_1-j)(v_1-iw_1)$, with ${\alpha_1}^2>1$, admits a polar form $z_1=\rho_1 e^{i\Omega_1}e^{j\Sigma_1}$, with $\rho_1$, $\Omega_1$, and $ \Sigma_1$ real parameters; hence we have the correspondence 
\begin{equation}
\tanh\Sigma_1 = -1/\alpha_1, \quad
\tan\Omega_1 = -w_1/v_1, \quad {\rho _1}^2= (\alpha^2_1-1)(v_1^2+w_1^2);
\label{pdpartial}
\end{equation}
however $z_1e^{i\theta}e^{j\chi}=\rho_1 e^{i\Omega_1+i\theta}e^{j\Sigma_1+j\chi}$ has also a modulus positive definite, which allows to find the general form of the expression (\ref{pd}) for $\alpha^2>1$,
\begin{equation}
z=[\alpha \cosh\chi-\sinh\chi-j(\cosh\chi-\alpha \sinh \chi)][v \cos\theta+w \sin \theta+i(v \sin \theta - w \cos \theta)];
\label{pdgeneral}
\end{equation}
similarly for a number of the form $z_{0}=(\alpha_0-j)(v_0-iw_0)$, with ${\alpha_0}^2<1$, admits the polar form $z_0=j\rho_0 e^{i\Omega_0}e^{j\Sigma_0}$, with $\rho_0$, $\Omega_0$, and $ \Sigma_0$ real parameters, and hence $z_0\overline{z_0}=-{\rho_0}^2$ , with the correspondence 
\begin{equation}
\tanh\Sigma_0 = -\alpha_0, \quad
\tan\Omega_0 = -w_0/v_0, \quad {\rho _0}^2= (1-\alpha^2_0)(v_0^2+w_0^2);
\label{pdpar}
\end{equation}
the expression (\ref{pdgeneral}) also works as the generalization for this case. Note that the hyperbolic phase does not affect the components $(v,w)$ of the restricted hypercomplex number, and hence does not correspond properly to a hyperbolic rotation. Similarly the choice $x=\beta y$, and $v=\beta w$, leads to a number of the form $z=(\beta+i)(y+jw)$, with norm $z\overline{z}=(\beta^2+1)(y^2-w^2)$, with a legitimate hyperbolic rotation and a spurious circular rotation. We develop now the case for a hypercomplex number described by two real quantities and with a norm invariant under a legitimate full phase $e^{i\theta} e^{j\chi}$.

The appropriate identification is $x=\gamma w$ and $y=\gamma v$, with $\gamma$ a real parameter, that reduces the expression (\ref{ring}) to
\begin{equation}
     z= (\gamma +ij)w+(i\gamma +j)v , \quad \overline{z} = (\gamma +ij)w-(i\gamma +j)v,\quad  z\overline{z} = (\gamma^{2}-1)(v^{2}+w^{2}) + 2ij\gamma (w^{2}-v^{2});
     \label{correct}
\end{equation}
hence, the norm is invariant under the interchange
of the field $v\leftrightarrow w$, and simultaneously the change $\gamma\rightarrow-\gamma$.
The effect of a combined circular and hyperbolic rotation is, 
\begin{eqnarray}
     e^{i\theta} e^{j\chi} z \!\! & = & \!\! (\gamma \cos\theta\cosh\chi - \sin\theta\sinh\chi)w + (\cos\theta\sinh\chi - \gamma\sin\theta\cosh\chi)v \nonumber \\
     \!\! & & \!\! +i[(\gamma\cos\theta\cosh\chi + \sin\theta\sinh\chi)v + (\gamma\sin\theta\cosh\chi + \cos\theta\sinh\chi)w] \nonumber \\
     \!\! & & \!\! +j[(\cos\theta\cosh\chi - \gamma\sin\theta\sinh\chi)v + (\gamma\cos\theta\sinh\chi - \sin\theta\cosh\chi)w] \nonumber \\
     \!\! & & \!\! +ij[(\sin\theta\cosh\chi + \gamma\cos\theta\sinh\chi)v + (\cos\theta\cosh\chi + \gamma\sin\theta\sinh\chi)w];
     \label{combined}
\end{eqnarray}
if $\gamma=0$, then $z=j(v+iw)$,  which is essentially an ordinary complex number; however, if $\gamma\neq 0$, then, as opposed to the cases  previously considered, the invariant norm (\ref{correct}) contains necessarily a $ij$-hybrid term, and it can not be a purely real quantity. However, the norm can be a purely hybrid quantity by choosing $\gamma^{2}=1$; note that in this case the number (\ref{correct}) can be reduced to a number proportional to a purely hyperbolic one, and the circular rotation is spurious. On the the hand, the case $\gamma^{2}\neq 1$ leads in its turn to two different cases, with $\gamma^{2}>1$, and $\gamma^{2}<1$. These algebraically different cases will lead to physically different scenarios, which will be considered below.

Furthermore, note that the first term proportional to $w$ in the expression (\ref{correct}) does not change under conjugation, behaving as the real part of an ordinary complex number; similarly the term proportional to $v$ will have a global change under conjugation, behaving as the imaginary part of an ordinary complex number; thus, one can reverse the expression (\ref{correct}) as
\begin{equation}
     w= \frac{\gamma -ij}{2(\gamma^{2}+1)} (z+\overline{z}), \qquad v= \frac{j-i\gamma}{2(\gamma^{2}+1)} (z-\overline{z}),
     \label{reverse}
\end{equation}
where we have used the inverse expressions,
\begin{equation}
     (\gamma +ij)^{-1} = \frac{\gamma -ij}{\gamma^{2}+1}, \qquad (i\gamma +j)^{-1} = \frac{j-i\gamma}{\gamma^{2}+1},
     \label{inverse}
\end{equation}
which are not singular in spite of belonging to a ring.

Just as any ``not-null'' hyperbolic number can be brought into polar form $\rho e^{j\chi}$, where $\rho\in{\cal R}$ or $\rho\in j\cdot{\cal R}$ depending on whether the norm of the number is strictly positive or strictly negative, respectively, any  number of the general form (\ref{ring}), with $|z|\neq 0$ can be written as
\begin{equation}
	z=\rho e^{i\theta}e^{j\chi} ,
\end{equation}
where $\theta\in (0,2\pi ]$, $\chi\in{\cal R}$ and, in general, $\rho$ must be a \emph{Hermitian} number, i.e.,
\begin{equation}
	\rho=\rho_R+ij\rho_H .
\end{equation}
Equating both expressions (Cartesian and polar) for $z$ we obtain the relations
\begin{eqnarray}
	x=\rho_R\cos\theta\cosh\chi-\rho_H\sin\theta\sinh\chi ,\quad 
	y=\rho_R\sin\theta\cosh\chi-\rho_H\cos\theta\sinh\chi ,\nonumber \\
	v=\rho_R\cos\theta\sinh\chi-\rho_H\sin\theta\cosh\chi ,\quad	w=\rho_R\sin\theta\sinh\chi-\rho_H\cos\theta\cosh\chi ;
\end{eqnarray}
which can in principle be inverted to obtain $\rho_R$, $\rho_H$, $\theta$ and $\chi$ in terms of $x$, $y$, $v$ and $w$. The explicit expressions turn out to be relatively complicated, but we can straightforwardly arrive to the following implicit formulas:
\begin{eqnarray}
	\rho_R^2-\rho_H^2=x^2+y^2-v^2-w^2 , \quad
	\rho_R\rho_H=xw-yv , \nonumber\\
	\frac{1}{2}(\rho_R^2+\rho_H^2)\sinh 2\chi =xv+yw ,\quad
	\frac{1}{2}(\rho_R^2+\rho_H^2)\sin 2\theta =xy+vw;
\end{eqnarray}
that is, given the Cartesian components of $z$, one can use the first two equations to obtain $\rho_R$ and $\rho_H$ in terms of them, and then insert those values into the latter two to get the phases $\theta$ and $\chi$.

In the hypercomplex formulation developed, the real objects such as Lagrangians, vector fields, masses, and coupling parameters will be generalized to Hermitian objects, encoding two real objects. The four real components of a hypercomplex field have been identified to each other by using a real $\gamma$-parameter, leading to two real effective variables; hence, the new formulation is constructed as a $\gamma$-deformation along a non-compact direction defined by the new complex unit. As an effect of the $\gamma$-deformation, the traditional Mexican hat potential will be hallowed out in two points in the valley that defines the degenerate vacuum; such two points represent the new vacuum states. In a limit case, a purely hyperbolic version of the Mexican hat potential will be obtained.

\section{ Hypercomplex version of the classical model $\lambda\phi^{4}$: global symmetries} 
\label{lambda4}
The theory ``$\lambda\phi^4$" is not only a pedagogical model; for example the potential $\lambda (H^{\dagger}H)^2$ has been considered recently for supporting the idea of a quantum origin of the Higgs potential and the electroweak scale \cite{hill}; specifically 
the renormalization group formalism for the corresponding Coleman-Weinberg potential obtained by radiative corrections is developed.
Hence, the Coleman-Weinberg symmetry breaking can be understood in terms of the running of the coupling constants. However, although the present formulation is far in spirit from the Coleman-Weinberg dynamical symmetry breaking scheme, the hypercomplex deformation of the $U(1)$ $\lambda\phi^4$ theory will be understood, in certain sense, in terms of the running of the coupling constants, as functions of the parameter $\gamma$, which has up to this point, an algebraic origin.

Now we shall show that, properly analytically continued, the $\lambda\phi^{4}_{d}$ theory for a massive (real) scalar field $\phi$, in a $d$-dimensional flat background, can be re-formulated on the hypercomplex space; the conventional theory is described by
\begin{equation}
     {\cal L} (\phi)= \int dx^{d} \Big(\frac{1}{2} \partial^{i} \phi \cdot \partial_{i} \phi - V(\phi)\Big), \qquad V(\phi) = \frac{1}{2} a m^{2} \phi^{2} + \frac{\lambda}{4!} \phi^{4},
     \label{lag}
\end{equation}
where  $m^2>0$, and the self-interacting constant $\lambda$ is assumed to be positive in order to have the energy bounded from below; the action in invariant under the action of the cyclic group $Z_{2}=\{+1,-1\}$, manifested though the discrete symmetry 
$\phi \rightarrow -\phi$.
The unbroken exact symmetry scenario requires $a=1$, and the vacuum manifold corresponds to a single point, without homotopy constraints, and hence without possible topological defects. Furthermore,
the spontaneous symmetry breaking scenario requires $a=-1$, and the vacuum manifold corresponds to the $0$-sphere, $S^{0} \sim Z_{2}$, with the non-trivial homotopy constraint  $\pi_{0}=2$, and hence admitting {\it the domain walls} as possible topological defects.
\begin{figure}[H]
  \begin{center}
    \includegraphics[width=.35\textwidth]{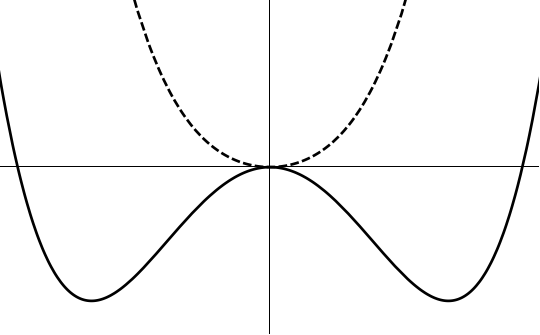}
  \caption{The dashed-potential shows the profile for the cases $m\geqslant 0$, and the continuous potential shows the case $m<0$; in the first case the vacuum-manifold corresponds to a single point, thus $\pi_{0}=1, \pi_{n\geqslant 1}=\{0\}$, and in the second case, it corresponds to  two points (the $0$-sphere), with $\pi_{0}=2, \pi_{n\geqslant 1}=\{0\}$.}  
   \label{uusual}
  \end{center}
\end{figure}
The hypercomplex version is based on the expressions (\ref{correct}), and (\ref{combined}), and hence the Lagrangian can be re-interpreted in terms of the two hyper-complexified fields variables $(\psi,\overline{\psi})$, with 
\begin{equation}
 \psi\overline{\psi}=(\gamma^2-1)(v^2+w^2)+2ij\gamma(w^2-v^2),
 \label{thenorm}
 \end{equation}
which is invariant under global phases $e^{i\theta}e^{j\chi}$, and under the discrete transformation $(v\leftrightarrow w, \gamma\rightarrow -\gamma)$;
\begin{equation}
     {\cal L}(\psi,\overline{\psi}) = \int dx^{d} \Big[\frac{1}{2} \partial^{i} \psi \cdot \partial_{i}\overline{\psi} - V(\psi ,\overline{\psi})\Big], \qquad V(\psi ,\overline{\psi}) = \frac{a}{2} m^{2} \psi\overline{\psi} + \frac{\lambda}{4!} \psi^{2}\overline{\psi}^{2},
     \label{complexlag}
\end{equation}
which is {\it Hermitian} and a non-analytical function on $\psi$, and hence can attain relative minimums and/or maximums; we consider that the square mass is also {\it Hermitian}, with real and hybrid parts $m^2\equiv m^2_{R}+ijm^2_{H}$; $a=\pm 1$, and similarly we consider that $\lambda\equiv \lambda_{R}+ij\lambda_{H}$, with $(m^2_{R}, m^2_{H}, \lambda_{R}, \lambda_{H}) $ real parameters. 
The potential can be written explicitly in terms of its real and hybrid parts as $V=V_{R}+ijV_{H}$,
\begin{eqnarray}
     V _{R}  \!\! & = & \!\!  a\Big(\frac{\gamma^{2}-1}{2} m^{2}_{R} + \gamma m^{2}_{H}\Big)v^{2} + a\Big(\frac{\gamma^{2}-1}{2} m^{2}_{R} - \gamma m^{2}_{H}\Big) w^{2}\nonumber\\ 
               \!\! & + & \!\!   \frac{\lambda_{R}}{6} \Big[ \frac{(\gamma^{2}-1)^{2}}{4} (v^{2}+w^{2})^{2} - \gamma^{2} (v^{2}-w^{2})^{2}\Big] -\frac{\lambda_{H}}{6}
    \gamma(\gamma^2-1)(w^4-v^4); 
     \label{VR} \\
   V_{H}  \!\! & = & \!\!  a\Big(\frac{\gamma^{2}-1}{2} m^{2}_{H} -\gamma m^{2}_{R}\Big) v^{2} + a \Big(\frac{\gamma^{2}-1}{2} m^{2}_{H} + \gamma m^{2}_{R}\Big)w^{2} + \frac{\gamma\lambda_{R}}{6} (\gamma^{2}-1) (w^{4}-v^{4})\nonumber \\
   \!\! & +& \!\! \frac{\lambda_{H}}{6} \Big[ \frac{(\gamma^{2}-1)^{2}}{4} (v^{2}+w^{2})^{2} - \gamma^{2} (v^{2}-w^{2})^{2}\Big];
     \label{potentialVH}
\end{eqnarray}
one can map the potentials $V_{R}$ and $V_{H}$ to each other,  by the discrete transformations 
\begin{equation}
\gamma \rightarrow -\gamma,\quad 
(\lambda_{R},\lambda_{H})\rightarrow (\lambda_{H},\lambda_{R}), \quad m_{R}\rightarrow m_{H}.
\label{discrete}
\end{equation}   
The vacuum is defined as usual by the stationary points constraint,
\begin{equation}
    \frac{\partial V}{\partial\psi_{0}} = \overline{\psi_{0}} [am^{2} + \frac{\lambda}{6} \psi_{0} \overline{\psi_{0}}] =0;
    \label{vacuum}
\end{equation}
which can be expressed explicitly is terms of real fields $(v_{0}, w_{0})$ and real parameters $(m_{R}, m_{H})$; the zero-energy point for $V_{R}$ and $V_{H}$ is described by
\begin{eqnarray}
      \Big( v_{0}=0,  w_{0}=0\Big); \label{zerozero}\\ 
       V_{R}=0;
      \quad ({\rm det} {\cal H})_{ (V_{R})}=4\Big[
     \big(\frac{\gamma^2-1}{2}\big)^2m^4_{R}-\gamma^2m^4_{H}\Big]; \nonumber\\
     \frac{\partial^2V_{R}}{\partial v^2_0}=2a\Big(\frac{\gamma^{2}-1}{2} m^{2}_{R} + \gamma m^{2}_{H}\Big); \quad \frac{\partial^2V_{R}}{\partial w^2_0}=2a\Big(\frac{\gamma^{2}-1}{2} m^{2}_{R} - \gamma m^{2}_{H}\Big); \label{falsev}\\
   \qquad V_{H} =0; \qquad \det {\cal H}_{ (V_{H})} = 4\Big[ \left(\frac{\gamma^{2}-1}{2}\right)^{2} m^{4}_{H} - \gamma^{2} m^{4}_{R}\Big]; \nonumber \\
     \frac{\partial^{2}V_{H}}{\partial v^{2}_{0}} =  2a \left(\frac{\gamma^{2}-1}{2} m^{2}_{H} - \gamma m^{2}_{R}\right), \qquad \frac{\partial^{2}V_{H}}{\partial w^{2}_{0}} = 2a \left(\frac{\gamma^{2}-1}{2} m^{2}_{H} + \gamma m^{2}_{R}\right); 
\label{falsevv}      
\end{eqnarray} 
where we have displayed the second order derivatives, and the determinant of the Hessian matrix; 
likewise,  the other stationary points for $V_{R}$ and $V_{H}$ related with the condition $ am^{2} + \frac{\lambda}{6} \psi_{0} \overline{\psi_{0}}=0$, read
\begin{eqnarray}
(1-\gamma^2)(\lambda_R^2+\lambda^2_H)(v_0^2+w_0^2)=6a(\lambda_R m_R^2+\lambda_H m^2_H);\label{circle}\\
\gamma(\lambda_R^2+\lambda^2_H)(v_0^2-w_0^2)=3a(\lambda_R m_H^2-\lambda_H m^2_R);\label{hyperbolae}
\end{eqnarray}which can be solved to favor of the fields $(v_0,w_0)$:
\begin{eqnarray}
v_{0}^2=\frac{3}{2a} \frac{1}{\lambda_{R}^2+\lambda_{H}^2}\frac{1}{\gamma(1-\gamma^2)}\Big\{\Big[ (\gamma^2-1)\lambda_{H}+2\gamma\lambda_{R}             \Big]m_{R}^2+ \Big[ (1-\gamma^2)\lambda_{R}+2\gamma\lambda_{H}  \Big]m_{H}^2  \Big\},\nonumber \\
w_{0}^2=\frac{3}{2a} \frac{1}{\lambda_{R}^2+\lambda_{H}^2}\frac{1}{\gamma(1-\gamma^2)}\Big\{\Big[ (1-\gamma^2)\lambda_{H}+2\gamma\lambda_{R}             \Big]m_{R}^2-   \Big[ (1-\gamma^2)\lambda_{R}-2\gamma\lambda_{H}  \Big]m_{H}^2  \Big\},
       \label{vacuumreal}\\ 
     V_{R}=\frac{3}{2}\frac{\lambda_{R}(m^4_{H}-m^4_{R})-2\lambda_{H}m^2_{H}m^2_{R}}{\lambda^2_{R}+\lambda^2_{H}}; \label{vacuumreal1}\\
      \frac{\partial^2V_{R}}{\partial v^2_0}=\frac{1}{3}\Big[(\gamma^4-6\gamma^2+1)\lambda _{R}+ 4\gamma(\gamma^2-1)\lambda_{H}\Big]v^2_{0}; \quad \frac{\partial^2V_{R}}{\partial w^2_0}=\frac{1}{3}\Big[(\gamma^4-6\gamma^2+1)\lambda _{R}- 4\gamma(\gamma^2-1)\lambda_{H}\Big]w^2_{0};\nonumber \\
      \Big(\frac{\partial^2V_{R}}{\partial v_0 \partial w_0}\Big)^2= \Big[  \frac{\lambda_{R}}{3}(\gamma^2+1)^2 v_{0}w_{0} \Big]^2;            
      \quad ({\rm det} {\cal H})_{ (V_{R})}=-\Big[ \frac{4\gamma(\gamma^2-1)}{3}\Big]^2(\lambda^2_{R}+\lambda^2_{H}) v^2_{0}w^2_{0} ;\label{vacuumreal2}\\
      V_{H}=\frac{3}{2}\frac{\lambda_{H}(m^4_{R}-m^4_{H})-2\lambda_{R}m^2_{H}m^2_{R}}{\lambda^2_{R}+\lambda^2_{H}}.
          \label{vacuumreal11}
          \end{eqnarray}
 Considering that $V_{H}$ is obtained from $V_{R}$ by mean of the transformations (\ref{discrete}), the second order derivatives expressions for $V_{H}$ can be obtained directly from the Eqs. (\ref{vacuumreal2}); in particular $(\det {\cal H})_{V_H} = (\det {\cal H})_{V_R}$, which is strictly negative, according to the expression (\ref{vacuumreal2}), and thus the points (\ref{vacuumreal}) are in general saddle points for both potentials, at least that $\det {\cal H} =0$, and then anything is possible. For this purpose, one can fix to zero one of the vacuum expectation values, $v_{0}$, or $w_{0}$, which in fact will be required by spontaneous symmetry breaking; the choice $w_{0}=0$ leads to the following simplification
\begin{eqnarray}
     w_{0}=0, \quad \rightarrow \quad \frac{m^{2}_{H}}{m^{2}_{R}} \!\! & = & \!\! \frac{(1-\gamma^{2})\lambda_{H}+ 2\gamma\lambda_{R}}{(1-\gamma^{2})\lambda_{R} -2\gamma\lambda_{H}}, \label{masscocient} \\
     v^{2}_{0} \!\! & = & \!\! \frac{6am^{2}_{R}}{(1-\gamma^{2})\lambda_{R} -2\gamma\lambda_{H}}; \label{expvalue}
\end{eqnarray}
fortunately, these conditions will induce two global minimums for both potentials $V_{R}$ and $V_{H}$ in the case with $\gamma^{2} \neq 1$, and with a non-zero vacuum expectation value for the field $v$; this case will be analyzed in detail in Section (\ref{gammano1}). Alternatively  one has the choice   
  \begin{eqnarray}
     v_{0}=0, \quad \rightarrow \quad \frac{m^{2}_{H}}{m^{2}_{R}} \!\! & = & \!\! \frac{(1-\gamma^{2})\lambda_{H}- 2\gamma\lambda_{R}}{(1-\gamma^{2})\lambda_{R} +2\gamma\lambda_{H}}, \label{masscocient2} \\
     w^{2}_{0} \!\! & = & \!\! \frac{6am^{2}_{R}}{(1-\gamma^{2})\lambda_{R} +2\gamma\lambda_{H}}; \label{expvalue2}
\end{eqnarray}which can be obtained from (\ref{masscocient}), and (\ref{expvalue}) by the change $\gamma\rightarrow -\gamma$; this case will be developed in section (\ref{nogamma11}).
                 
Now we expand the theory around $\psi_{0}$, using only the degenerate vacuum constraint $\lambda\psi_{0}\overline{\psi}_{0} = -6am^{2}$,
\begin{eqnarray}
     V(\psi+\psi_{0}, \overline{\psi}+\overline{\psi_{0}}) \!\! & = & \!\! \frac{am^{2}}{2} (\psi + \psi_{0})(\overline{\psi} + \overline{\psi_{0}}) + \frac{\lambda}{4!} (\psi + \psi_{0})^{2}(\overline{\psi}+\overline{\psi_{0}})^{2} \nonumber \\
     \!\! & = & \!\!- \frac{am^{2}}{2} \psi\overline{\psi} + \frac{\lambda}{4!} (\psi\overline{\psi})^{2} + \frac{\lambda}{4!} (\overline{\psi}^{2}_{0}\psi^{2} + \psi^{2}_{0}\overline{\psi}^{2}) + \frac{\lambda}{12} \psi\overline{\psi} (\psi_{0}\overline{\psi}+ \overline{\psi}_{0}\psi),
     \label{neve}
\end{eqnarray}
note that we have not chosen yet a definite vacuum, but the expansion around a nonzero ground state value for the field leads to a change of sign in the mass term. Furthermore, the first two terms in the expression (\ref{neve}) have already the canonical form since depend on the norm $\psi\overline{\psi}$ given in (\ref{thenorm}). The third and fourth terms correspond to the quadratic and cubic terms in the fields $(v,w)$; such terms are not in the canonical form (due to the presence of mixed terms of the form $vw$) and do not depend on the modulus $\psi_{0}\overline{\psi}_{0}$; explicitly we have,
\begin{eqnarray}
     \overline{\psi}^{2}_{0} \psi^{2} + \psi^{2}_{0}\overline{\psi}^{2} \!\! & = & \!\! 2\cosh 2(\chi_{0}-\chi) \cos 2(\theta_{0}-\theta)  \Big\{ (\gamma^{4}+1)(w^{2}_{0}-v^{2}_{0})(w^{2}-v^{2}) -2\gamma^{2} [(v^{2}_{0}+3w^{2}_{0})w^{2} + (3v^{2}_{0}+w^{2}_{0})v^{2}]\Big\} \nonumber \\
     \!\! & & \!\! +8\gamma \sinh 2(\chi_{0}-\chi) \Big\{ (\gamma^{2}-1) \sin 2(\theta_{0}-\theta )(v^{2}_{0}v^{2}-w^{2}_{0}w^{2}) - (\gamma^{2}+1) \cos 2(\theta_{0}-\theta)v_{0}w_{0}(v^{2}+w^{2})\Big\} \nonumber \\
     \!\! & & \!\! +4 (\gamma^{4}-1) \cosh 2(\chi_{0}-\chi) \sin 2(\theta_{0}-\theta) v_{0}w_{0}(v^{2}-w^{2}) \nonumber \\
     \!\! & & \!\! +8 (\gamma^{2}+1) \cos 2(\theta_{0}-\theta) \Big[(\gamma^{2}+1) v_{0}w_{0} \cosh 2(\chi_{0}-\chi) + \gamma (v^{2}_{0}+w^{2}_{0}) \sinh 2(\chi_{0}-\chi)\Big]\underbrace {vw }\nonumber \\
     \!\! & & \!\! -4 (\gamma^{4}-1) (v^{2}_{0}-w^{2}_{0}) \sinh 2(\chi_{0}-\chi) \sin 2(\theta_{0}-\theta) \cdot \underbrace{vw }\nonumber \\
     \!\! & + & \!\! 2ij \Big\{ 4\gamma (\gamma^{2}-1) \cosh 2(\chi_{0}-\chi) \cdot \cos 2(\theta_{0}-\theta) (w^{2}_{0}w^{2} - v^{2}_{0}v^{2}) \nonumber \\
     \!\! & & \!\! + \sinh 2(\chi_{0}-\chi) \sin 2(\theta_{0}-\theta) \Big[(\gamma^{4}+1)(w^{2}_{0}-v^{2}_{0}) (w^{2}-v^{2}) - 2\gamma^{2} (v^{2}_{0}+w^{2}_{0}) (v^{2}+w^{2}) \nonumber \\
     \!\! & & \!\!  -4 \gamma^{2} (v^{2}_{0}v^{2} +w^{2}_{0}w^{2})\Big] \nonumber \\
     \!\! & & \!\! +2 (\gamma^{4}-1) \sinh 2(\chi_{0}-\chi) \cos 2(\theta_{0}-\theta) v_{0}w_{0} (w^{2}-v^{2}) \nonumber \\
     \!\! & & \!\! -4 \gamma(\gamma^{2}+1) \cosh 2(\chi_{0}-\chi) \sin 2(\theta_{0}-\theta) v_{0}w_{0} (w^{2}+v^{2}) \Big\} \nonumber \\
     \!\! & + & \!\! 4ij (\gamma^{2}+1) \sinh 2(\chi_{0}-\chi) \Big[2v_{0}w_{0} \sin 2(\theta_{0}-\theta) - (\gamma^{2}-1)(w^{2}_{0}-v^{2}_{0}) \cos 2(\theta_{0}-\theta)\Big]\underbrace{ vw }\nonumber \\
     \!\! & + & \!\! 8 \gamma ij(\gamma^{2}+1) \cosh 2(\chi_{0}-\chi) \sin 2(\theta_{0}-\theta) (v^{2}_{0}+w^{2}_{0})\underbrace{vw};
     \label{quacub1}
\end{eqnarray}
\begin{eqnarray}
     \psi\overline{\psi}(\psi_{0}\overline{\psi} + \overline{\psi}_{0}\psi) \!\! & = & \!\! 2\Big[ (\gamma^{2}-1)(v^{2}+w^{2}) + 2\gamma ij(w^{2}-v^{2})\Big] \cdot \nonumber \\
     \!\! & & \!\! \Big\{(\gamma^{2}-1) \cosh (\chi_{0}-\chi) \cos (\theta_{0}-\theta) (w_{0}w+v_{0}v) - 2\gamma \sinh (\chi_{0}-\chi) \sin (\theta_{0}-\theta) (w_{0}w-v_{0}v) \nonumber \\
     \!\! & & \!\! -(\gamma^{4}+1) \cosh (\chi_{0}-\chi) \sin (\theta_{0}-\theta) (v_{0}w-w_{0}v) \nonumber \\
     \!\! & + & \!\! ij \Big[(\gamma^{2}-1) \sinh (\chi_{0}-\chi) \sin (\theta_{0}-\theta) (w_{0}w+v_{0}v) + 2\gamma \cosh (\chi_{0}-\chi) \cos (\theta_{0}-\theta) \cdot (w_{0}w -v_{0}v) \nonumber \\
     \!\! & & \!\!  + (\gamma^{2}+1) \sinh (\chi_{0}-\chi) \cos (\theta_{0}-\theta) (v_{0}w-w_{0}v)\Big]\Big\};
     \label{quacub2}
\end{eqnarray}     
where the fields have the general form (\ref{combined}), with $\psi_{0} = \psi_{0}(\gamma, v_{0},w_{0}; \theta_{0},\chi_{0})$ and $\psi= \psi(\gamma,v,w;\theta,\chi)$; note the presence of $ij$-hybrid terms in the expressions (\ref{quacub1}), and (\ref{quacub2}). Furthermore, the presence of mixed terms $vw$ (both real and hybrid) in the quadratic form (\ref{quacub1}) prevents us from determining the masses of $v$ and $w$; note also that in relation to the same mixed terms, one must exploit the freedom of choosing the circular and hyperbolic parameters, and v.e.v. $(v_{0}, w_{0})$, in order to obtain the simultaneous vanishing of the real and hybrid terms of the form $vw$.

Finally the equations of motion are given by
\begin{equation}
     \Big[\Box + \frac{\lambda}{6} \left(\psi\overline{\psi}- \frac{6am^{2}}{\lambda}\right)\Big]\psi =0, \quad \Box = \partial^{2}t - \nabla^{2},
     \label{em}
\end{equation}
with an energy momentum tensor given by
\begin{equation}
     2T_{ij} = \partial_{i}\psi \cdot \partial_{j} \psi - g_{ij} \Big[\frac{1}{2} \partial^{\kappa}\psi\cdot\partial_{\kappa}\psi - \frac{\lambda}{4!} \left(\psi\cdot\overline{\psi} - \frac{6am^{2}}{\lambda}\right)^{2}\Big];
     \label{emt}
\end{equation}
in particular, the energy density reads
\begin{equation}
     {\cal E} = 2T_{00}=\frac{1}{2}|\partial_{ t} \psi|^{2} + \frac{1}{2} \nabla\psi\cdot\nabla\overline{\psi} + \frac{\lambda}{4!} \Big(\psi\overline{\psi} - \frac{6am^{2}}{\lambda}\Big)^{2}.
     \label{energy}
\end{equation}
Eqs. (\ref{em}), (\ref{emt}), and (\ref{energy}) will allow to study the possible formation of global strings with finite energy in the section (\ref{topdef}).

\subsection{The case $\gamma^2=1$, $m^2_{R}=0$, $\lambda_H=0$: hyperbolic version of the mexican hat}
\label{gamma1}

The case $\gamma^2=1$ is special, since the norm (\ref{thenorm}) reduces to the hyperbolic part, a purely hybrid expression. In relation to the Eq. (\ref{circle}) that defines the degenerate vacuum, the restriction $\gamma^{2}=1$ implies more
restrictions on the right-hand side; a non-trivial  choice is $m_{R}=0$, and $\lambda_H=0$; then the hybrid component of the potential  $V_H$ (Eq. \ref{potentialVH}) vanishes, and its real part reduces to
\begin{equation}
     V = V_{R}= a\gamma m_{H}^{2} (v^{2}-w^{2}) - \frac{\lambda}{6} (v^{2}-w^{2})^{2}, \quad V_{H}=0;
     \label{nohybrid}
\end{equation}
additionally the minima are defined by the hyperbola (\ref{hyperbolae}), irrespective of the signs of the parameters $(a,\gamma ,\lambda_R=\lambda)$;
\begin{equation}
     a\gamma (v^{2}_{0}-w^{2}_{0}) = \frac{3}{\lambda} m^{2}_{H};
     \label{irres}
\end{equation}
the hyperbola with $a\gamma>0$ is the conjugate of that with $a\gamma<0$, and are related by a $\frac{\pi}{2}$-rotation in the plane $v-w$. In the appropriate field variables $(\Theta_{1}, \Theta_{2})$, the  hyperbola takes the form $\Big( \frac{\Theta_{1}}{\sqrt{|\frac{6m^{2}_H}{\lambda}}|} \Big)^{2} - \Big( \frac{\Theta_{2}}{\sqrt{|\frac{6m^{2}_H}{\lambda}}|} \Big)^{2} =1$, and thus is rectangular with excentricity $\sqrt{2}$, with the vertices localized at $\pm \sqrt{|\frac{6m^{2}_H}{\lambda}|}$, which coincide with the position of the two-minimums in the original Spontaneous Symmetry Breaking (SSB) scenario described in terms of the real field $\phi$ in figure 1. From the substitution of the expression (\ref{irres}) into the Eq. (\ref{nohybrid}), one obtains the energy of vacuum,
\begin{equation}
     V(v_{0},w_{0}) = \frac{3}{2} \frac{m^{4}_{H}}{\lambda},
     \label{energygamma1}
\end{equation}
For $\lambda <0$, the hyperbola corresponds to the global minima for the energy, and whose value corresponds to the value for the two lowest energy states in the original SSB scenario described in figure 1. For the stationary points described in Eq. (\ref{zerozero}) we have that
${\rm det}{\cal H}_{(V_{R})}(0,0)=-4m^4_H$, and thus the zero-energy point is a saddle point.
\begin{figure}[H]
  \begin{center}
    \includegraphics[width=.5\textwidth]{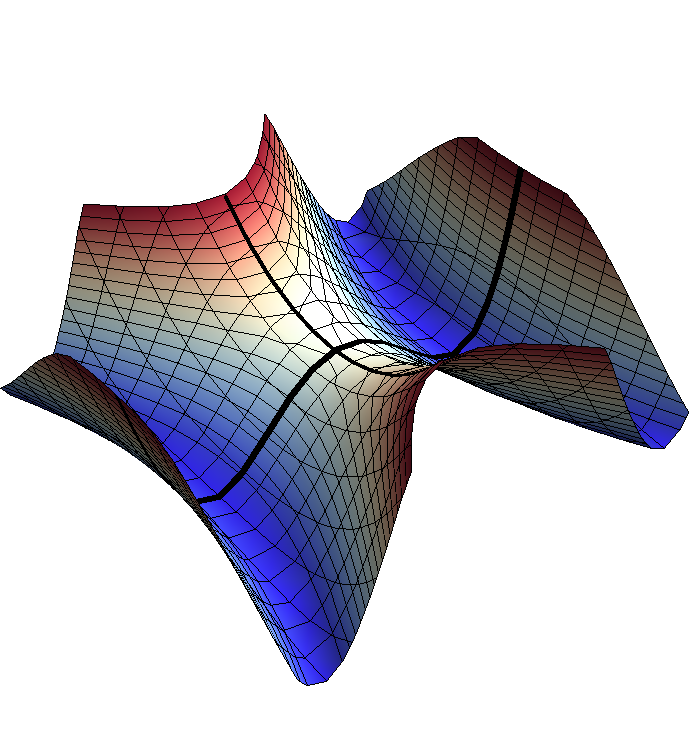}
  \caption{The potential (\ref{nohybrid}) for $\lambda<0$; the stable states are localized at the blue region, a hyperbola described by Eq. (\ref{irres}).}  
   \label{gamma}
  \end{center}
\end{figure}
In the figure \ref{gamma}, the potential  is not the Mexican hat potential, particularly in relation to the existence of two connected regions for the possible vacuum states, and that each region is non-compact, as opposed to the compact circle associated to the conventional U(1)-symmetry.
Let us see how the two scenarios described in the conventional Lagrangian (\ref{lag}) are contained in certain sense in this hypercomplex re-formulation; 
if the original Lagrangian corresponds to an Unbroken Exact Symmetry (UES) scenario with $L(\phi , a=1)$, then the potential $V(\phi , a=1)$ has a minimum (see figure \ref{uusual}), and is shown as a {\it bold-face} curve embedded in the figure \ref{gamma}; the hypercomplex extension of $\phi$ leads to a new potential, transforming the original stable minimum into a saddle point as the zero-energy point, and with the appearance of other stable minimums at the blue region. Similarly,  the usual SSB scenario with $V(\phi , a=-1)$ in figure (\ref{uusual})  with two minimums, is shown also as a {\it bold-face} curve in the figure \ref{gamma}. Thus, the two minimums have been extended to an infinite number of minimums in two disconnected regions at the blue regions, which correspond to the hyperbola in Eq. (\ref{irres}); the original maximum point of the potential $V(\phi , a=-1)$ is the saddle point for the new potential. Hence, the Lagrangian (\ref{complexlag}) is build originally on a saddle point (the crossing point of the {\it bold-face} curves), and the choice a stable vacuum in the blue region is mandatory, and we proceed now to the study of the realizations of the symmetry breakdown.

In this case the expressions (\ref{quacub1}), and  (\ref{quacub2}) reduce to
\begin{eqnarray}
     \overline{\psi}^{2}_{0} \psi^{2} + \psi^{2}_{0}\overline{\psi}^{2} \!\! & = & \!\! 4\cosh 2(\chi_{0}-\chi) \cos 2(\theta_{0}-\theta) \Big[\frac{3a\gamma}{\lambda} m^{2}_{H}(v^{2}-w^{2}) - (v^{2}_{0}+3w^{2}_{0})w^{2} - (3v^{2}_{0}+w^{2}_{0})v^{2}\Big] \nonumber \\
     \!\! & & \!\! -16\gamma\sinh 2(\chi_{0}-\chi) \cdot \cos 2(\theta_{0}-\theta ) v_{0}w_{0} (v^{2}+w^{2}) \nonumber \\
     \!\! & & \!\! + 16 \cos 2(\theta_{0}-\theta )\Big[ 2v_{0}w_{0}\cosh 2(\chi_{0}-\chi) + \gamma (v^{2}_{0}+w^{2}_{0})\sinh 2(\chi_{0}-\chi)\Big]\underbrace{vw} \nonumber \\
     \!\! & & \!\! +4ij \sinh 2(\chi_{0}-\chi) \sin 2(\theta_{0}-\theta ) \Big[\frac{3a\gamma}{\lambda} m^{2}_{H} (v^{2}-w^{2}) - (v^{2}_{0}+w^{2}_{0})(v^{2}+w^{2}) - 2(v^{2}_{0}v^{2}+w^{2}_{0}w^2)\Big] \nonumber \\
     \!\! & & \!\! +16\gamma ij\cosh 2(\chi_{0}-\chi) \sin 2(\theta_{0}-\theta ) \Big[(v^{2}_{0}+w^{2}_{0})vw - v_{0}w_{0}(v^{2}+w^{2})\Big] \nonumber \\
     \!\! & & \!\!  + 16ij \sinh 2(\chi_{0}-\chi) \sin 2(\theta_{0}-\theta ) v_{0}w_{0}\underbrace{vw}, \label{expandgamma1} \\
     \psi\overline{\psi}(\psi_{0}\overline{\psi} +\overline{\psi}_{0}\psi) \!\! & = & \!\! 8\gamma ij (w^{2}-v^{2}) \Big[ \gamma\sinh 2(\chi_{0}-\chi) \sin 2(\theta_{0}-\theta ) (v_{0}v-w_{0}w) \nonumber \\
     \!\! & & \!\! + \cosh 2(\chi_{0}-\chi) \sin 2(\theta_{0}-\theta ) (w_{0}v-v_{0}w) + ij\sinh 2(\chi_{0}-\chi) \cos 2(\theta_{0}-\theta ) (v_{0}w-w_{0}v) \nonumber \\
     \!\! & & \!\! + \gamma ij\cosh 2(\chi_{0}-\chi) \cos 2(\theta_{0}-\theta ) (w_{0}w-v_{0}v)\Big];
     \label{expandgamma11}
\end{eqnarray}     
We can see that only some mixed terms $vw$ (both real and hybrid), can be gauged away by fixing the hyperbolic parameters $\chi = \chi_{0}$; a particular choice, say $\chi_{0}=0$, leads to a break down of the symmetry $SO^{+}(1,1)$, remembering that in the case $\gamma^2=1$, $U(1)$ corresponds to a spurious rotation.  Even so the remanent $U(1)$ parameters must be fixed in Eq. (\ref{expandgamma1}) by demanding the vanishing of the remaining mixed term $v\cdot w$, for example, by fixing 
\begin{equation}
\theta=\theta_{0}
=0, \quad v_{0}=0,  \quad  w_{0}^{2} = -\frac{3a\gamma}{\lambda}m^{2}_{H};\quad
\rightarrow \frac{\lambda}{4!}(\overline{\psi}^{2}_{0} \psi^{2} + \psi^{2}_{0} \overline{\psi}^{2} )=  a\gamma m^{2}_{H} (w^{2}+v^{2}),
\label{fixing1}
\end{equation}
 with $a\gamma=1$, and $\lambda<0$; hence, taking into the account that $-\frac{am^2}{2}\psi\overline{\psi}=a\gamma m^2_{H}(w^2-v^2)$, the mass matrix determined by Eq. (\ref{neve})  for the Lagrangian (\ref{complexlag}) reads
\begin{eqnarray}
\left( \begin{array}{cc}
                 0 & 0 \\
                 0 & -2m^2_{H}w^2
                 \end{array} \right),
     \label{matrix1}
\end{eqnarray}  
 which corresponds to an massive ordinary scalar field $w$, and a massless field $v$. Conversely, if the field $v$ develops a nonzero vacuum expectation value with $ v_{0}^2 =\frac{3a\gamma}{\lambda}m^{2}_{H}$,  $a\gamma=-1$, and $w_{0}=0$, then we shall have a ordinary massive term for the field $v$, and the field $w$ is now massless; in this case the mass matrix reads
\begin{eqnarray}
\left( \begin{array}{cc}
                 -2m^2_{H}v^2 & 0 \\
                 0 & 0
                 \end{array} \right).
     \label{matrix2}
\end{eqnarray} 
If the spurious  $U(1)$  rotations are  broken for example with the choice (\ref{fixing1}), then the mixed terms $vw$ in the real part of the Eq. (\ref{quacub1}) vanish trivially, and the complete mass term in Eq. (\ref{neve}) reduces to
\begin{equation}
      -\frac{am^2}{2}\psi\overline{\psi} + \frac{\lambda}{4!}(\overline{\psi}^{2}_{0} \psi^{2} + \psi^{2}_{0} \overline{\psi}^{2} ) =  m^{2}_{H} \Big\{[\cosh 2(\chi-\chi_{0})-1] v^{2} + [1+\cosh 2(\chi-\chi_{0})]w^{2}-4\gamma\sinh2(\chi-\chi_{0})vw\Big\},
     \label{hypmass}
\end{equation} 
which is fully real, without $ij$-hybrid terms; similarly the cubic expression (\ref{expandgamma11}) is fully real under the choice (\ref{fixing1}). Therefore, in relation to the expression (\ref{hypmass}), whereas the hyperbolic rotation symmetry is no conveniently fixed, the masses of the fields $v$ and $w$ can not to be determined due to the presence of the mixed term $vw$. However, the choice $\chi_{0}=0$ fixes a point on the hyperbola, and the condition $\chi =0$ leads to the same mass matrix described previously in Eq.  (\ref{matrix1}), and similarly for the mass matrix in Eq.\ (\ref{matrix2}). 

In the spontaneous symmetry breaking scenarios described above, the hyperbolic parameters are always chosen with finite values, in similarity with the compact circular parameters; however, the hyperbolic parameters take in principle values on a non-compact interval, with $|\chi|<\infty$, and $|\chi_{0}|<\infty$.
Hence, the remanent $SO^{+}(1,1)$ symmetry in Eq. (\ref{hypmass}) can be  spontaneously broken by considering that $\chi_{0}\rightarrow +\infty$, and hence $\sinh\chi_{0}\approx\cosh\chi_{0}\approx \frac{e^{\chi_{0}}}{2}$, and for the hyperbolic functions in the expression (\ref{hypmass}) we have that
\begin{equation}
\cosh 2(\chi-\chi_{0})\approx \frac{e^{2\chi_{0}}}{2}[\sinh\chi-\cosh\chi]^2, \quad
\sinh2(\chi-\chi_{0})\approx -\frac{e^{2\chi_{0}}}{2}[\sinh\chi-\cosh\chi]^2,
 \label{hypmass2}
\end{equation} 
which are divergent, at least that $\chi\rightarrow +\infty$, and hence $\sinh\chi-\cosh\chi\approx 0$;  thus $\lim [\cosh 2(\chi-\chi_{0})] = 1$, and 
$\lim [\sinh 2(\chi-\chi_{0})] = 0$; therefore the mass matrix obtained from (\ref{hypmass}) coincides with that obtained previously with the choice $\chi=\chi_{0}=0$. Similarly, in the case of the another asymptotic limit $\chi_{0}\rightarrow -\infty$ we have that $-\sinh\chi_{0}\approx\cosh\chi_{0}\approx \frac{e^{-\chi_{0}}}{2}$, and the hyperbolic expressions depend now on $\sinh\chi+\cosh\chi$, which goes to zero in the limit $\chi\rightarrow -\infty$. 

The hyperbola has two connected regions, and hence $\pi_{0}=2$; each connected region is topolo\-gi\-cally equivalent to $R$, which has trivial homotopy groups, thus $\pi_{n}=0$, for $n\geq 1$; there are not topological defects associated with continuous symmetries, and only the {\it domains walls} are possible due to the non-triviality of $\pi_{0}$.

Since the structure of this theory is basically hyperbolic, it reduces in essence to the substitution of the ordinary imaginary unit $i$, by the new hyperbolic unit $j$;  as already mentioned, this simple substitution generates nontrivial results in diverse scenarios \cite{emam,emam1,perry,cortes,cortes1}. In the hyperbolic $``\lambda\phi^4"$ theory at hand, the topological defects such as strings, can not form; these defects are unavoidable in the breaking of an Abelian $U(1)$ symmetry. Hence, a phenomenological field theory based 
on a hyperbolic symmetry may be a way of evading the problem of the insignificant empirical evidence of cosmic strings, contrary to the predictions of many field theory models. The possible cosmological implications of these speculations are discussed in the concluding remarks.

Now we are going beyond the substitution of $i$ by $j$, and the incorporation of $j$ described previously in terms of a commutative ring will allow us to study the hyperbolic deformation of the mexican hat potential; the parameter $\gamma$ will govern the competition between the circular and hyperbolic contributions.

\subsection{The case $\gamma^2 \neq 1$, $(v_0\neq 0,w_0=0)$, and $\lambda_{R}=\lambda_{H}$: hyperbolic deformation of the Mexican hat}
\label{gammano1}
\begin{figure}[H]
  \begin{center}
  \includegraphics[width=.5\textwidth]{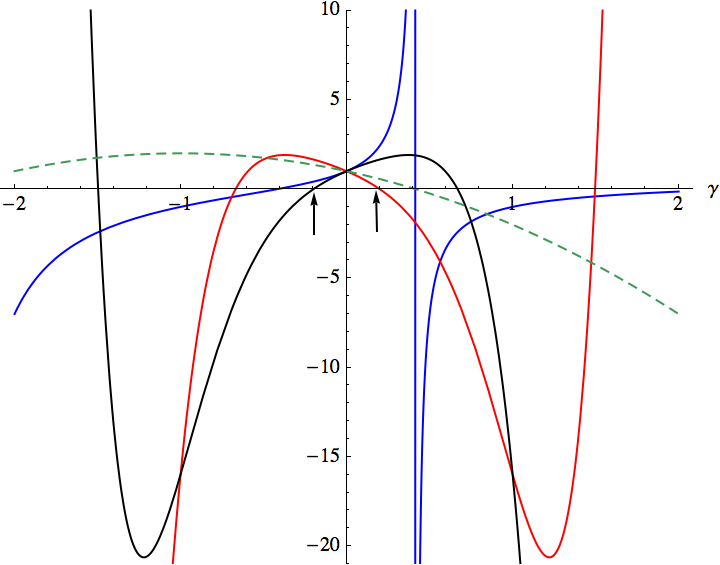}
\end{center}
\caption{The continuous red curve represents essentially $\det {\cal H}_{V_{R}}(0,0)$; the continuous black curve represents $\det {\cal H}_{V_{H}}(0,0)$.
The continuous blue curve represents the polynomial $(\gamma^2-2\gamma-1)/(\gamma^2+2\gamma^2-1)$, the ratio in  Eq. (\ref{positivequan}); the dashed curve represents the polynomial $(1-\gamma^2-2\gamma)$, Eq. (\ref{ineq4}); the arrows point out the interval where all polynomials are positive.}
\label{poly4}
\end{figure}

The restrictions $\gamma^{2} \neq 1$ and $\gamma \neq 0$ imply that the norm (\ref{thenorm}) will have nontrivial contributions from the {\it circular} and {\it hyperbolic} parts; the additional restriction $\lambda_{H} = \lambda_{R} \equiv \lambda$ will allow us to simplify the analysis; hence the expressions (\ref{masscocient}) and (\ref{expvalue}) for the ratio between the squared masses and squared expectation value, reduce to
\begin{equation}
     \frac{m^{2}_{H}}{m^{2}_{R}} = \frac{\gamma^{2}-2\gamma -1}{\gamma^{2}+2\gamma -1}, \quad v^{2}_{0} = \frac{6m^{2}_{R}}{a\lambda (1-\gamma^{2}-2\gamma)};
     \label{positivequan}
\end{equation}
the first equation above defines a positive quotient,  restricting  the values of $\gamma$ on the right-hand side;  similarly positivity on the lefht-hand side of the second equation implies the inequality
\begin{equation}
     a\lambda (1-\gamma^{2}-2\gamma )> 0.
     \label{ineq4}
\end{equation}
Other relevant quantities for inducing a stable vacuum are  $\det {\cal H}$ for the potentials at the zero energy-point, 
\begin{eqnarray}
     \det {\cal H}_{v_{R}} (v_{0}=0, w_{0}=0) \!\! & = & \!\! \frac{(\gamma^2+1)^2}{m^{2}_{R}(1-\gamma^{2}-2\gamma)^{2}} [\gamma^4+4\gamma^3-6\gamma^2-4\gamma+1], \label{zerostable1} \\
     \det {\cal H}_{v_{H}} (v_{0}=0, w_{0}=0) \!\! & = & \!\! \frac{(\gamma^2+1)^2}{m^{2}_{R}(1-\gamma^{2}-2\gamma)^{2}} [\gamma^4-4\gamma^3-6\gamma^2+4\gamma+1]; \label{zerostable2}
\end{eqnarray} 
if $\det {\cal H}>0$, the zero-energy point will be a minimum or maximum; if  $\det {\cal H}<0$ it will be a saddle point.
The polynomials of $\gamma$ in Eqs.\ (\ref{positivequan}), and (\ref{ineq4}), and those that determine the signs of $\det {\cal H}$ in Eqs.\   (\ref{zerostable1}), (\ref{zerostable2}), are shown in the figure (\ref{poly4}).

\begin{figure}[H]
\includegraphics[width=.53\textwidth]{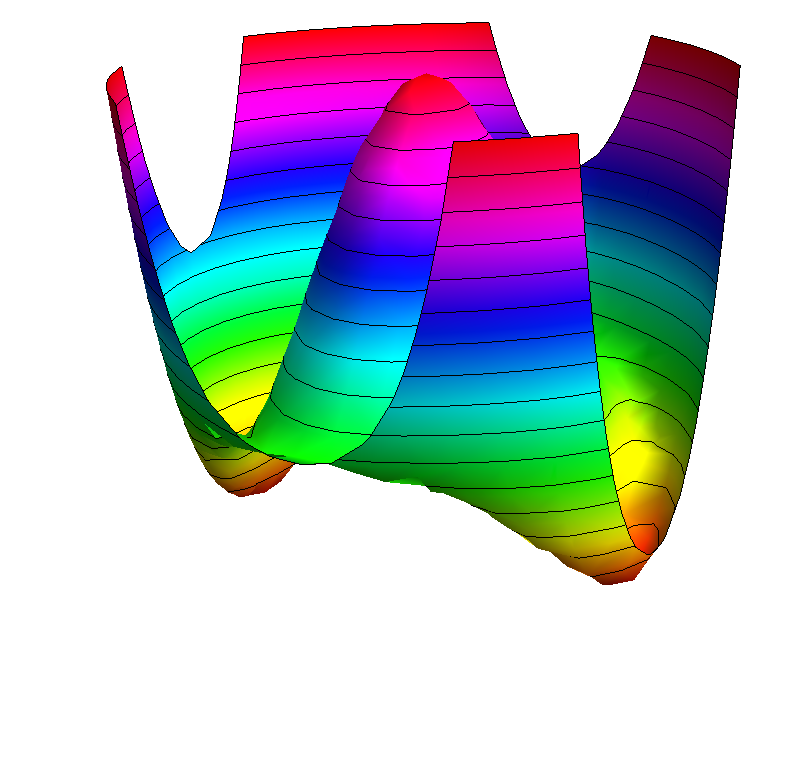}
  \includegraphics[width=.53\textwidth]{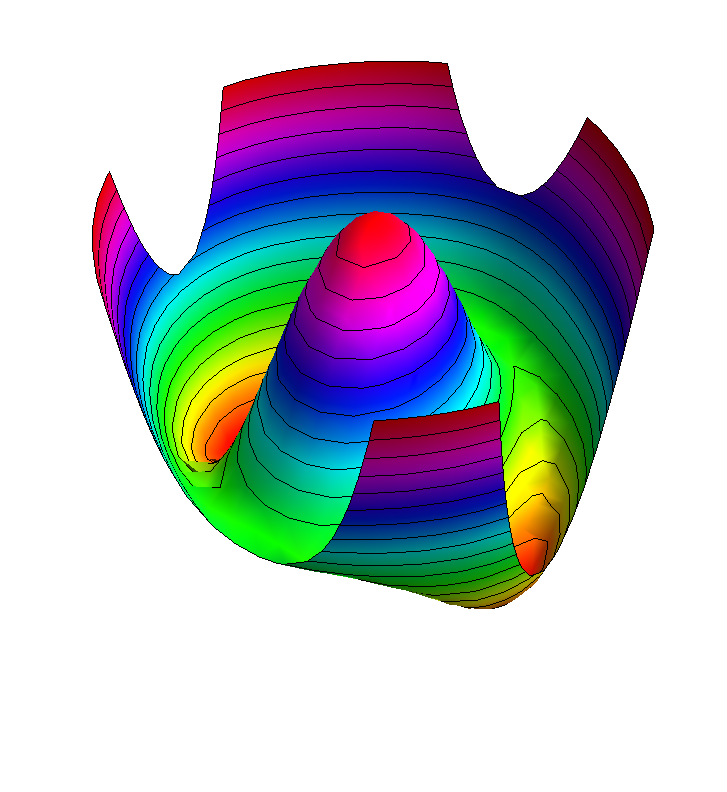}
\caption{The potentials $V_{R}/m^2_{R}$, and $V_{H}/m^2_{R}$  in the interval $(\gamma_{_{H}},-\gamma_{_{H}})$ have essentially the same qualitative aspect, which we show here from two different perspectives: the zero-energy point corresponds to the  central peak of the potential; the two minima are localized at the bottom in the two red regions:$(\pm v_0,0)$. The inequality (\ref{ineq44}) is satisfied with $a=1$ and $\lambda>0$; the choice $a=-1$ and $\lambda<0$ turns the potentials upside-down.}
\label{fig11}
\end{figure}

\begin{figure}[H]
  \begin{center}
  \includegraphics[width=.6\textwidth]{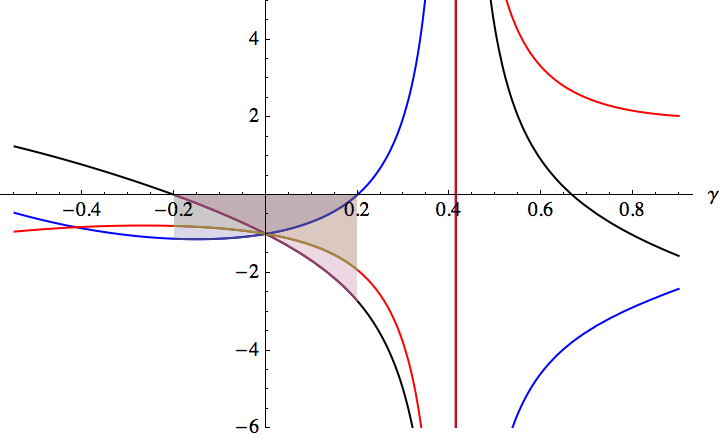}
\end{center}
\caption{The mass polynomial coefficients: the blue curve represents  $P^{v}_{R}$; the black curve represents $P^{v}_{H}$, and the red curve represents $P^{w}_{R}$. In the shadowed interval $(\gamma_{_{H}},-\gamma_{_{H}})$ all polynomials are ne\-ga\-ti\-ve, and $P^{v}_{R}(-\gamma_{_{H}})=0=P^{v}_{H}(\gamma_{_{H}})$, $P^{v}_{R}(\gamma_{_{H}})\approx -1.1252$ ,$P^{v}_{H}(-\gamma_{_{H}})\approx -2.7165$, and  all polynomials satisfy $P^{v}_{R}(0)=P^{v}_{H}(0)=P^{w}_{R}(0)=-1$.}
\label{polmass}
\end{figure}

In the figure (\ref{poly4}), the vertical blue asymptote represents one root of the polynomial $(\gamma^{2}+2\gamma-1)$, $\gamma=\sqrt{2}-1$; this singular point is out of the interval of interest. The arrow on the left-hand side points out a root of $\det {\cal H}_{V_{H}}(0,0)$, $\gamma_{_{H}}=1 + \sqrt{2} - \sqrt{2 (2 + \sqrt{2}})\approx -0.1989$; similarly on the right-hand side an arrow points out  a root of $\det {\cal H}_{V_{R}}(0,0)$  localized at $\gamma_{_{R}}=-\gamma_{_{H}}\approx 0.1989$; within this symmetric interval all polynomials are positive, and a stable vacuum will be induced for the potentials. In this interval, the inequality (\ref{ineq4}) implies that
\begin{equation}
a\lambda>0. \label{ineq44}
\end{equation} 
Under these simplifications the potentials will take the following form,
\begin{eqnarray}
V_{R}= \frac{a}{2}P^{v}_{R}m^2_{R}v^2+\frac{a}{2}P^{w}_{R}m^2_{R}w^2+ \frac{\lambda}{6} \Big[ \frac{(\gamma^{2}-1)^{2}}{4} (v^{2}+w^{2})^{2} - \gamma^{2} (v^{2}-w^{2})^{2} -
    \gamma(\gamma^2-1)(w^4-v^4)\Big]; 
     \label{VRS} \\
V_{H}= \frac{a}{2}P^{v}_{H}m^2_{R}v^2+\frac{a}{2}P^{w}_{H}m^2_{R}w^2+ \frac{\lambda}{6} \Big[ \frac{(\gamma^{2}-1)^{2}}{4} (v^{2}+w^{2})^{2} - \gamma^{2} (v^{2}-w^{2})^{2} +
    \gamma(\gamma^2-1)(w^4-v^4)\Big]; 
     \label{VHS}    
     \end{eqnarray} 
 where the polynomials $P$ are defined as
 \begin{eqnarray}
 P^{v}_{R}=\frac{\gamma^4+4\gamma^3-6\gamma^2-4\gamma+1}{\gamma^2+2\gamma-1}, \quad P^{w}_{R}= \frac{(\gamma^2+1)^2}{\gamma^2+2\gamma-1} ; \nonumber\\ 
  P^{v}_{H}=\frac{\gamma^4-4\gamma^3-6\gamma^2+4\gamma+1}{\gamma^2+2\gamma-1}, \qquad P^{w}_{H}=  P^{w}_{R}; \label{polymass}\end{eqnarray}    
the potentials (\ref{VRS}), and (\ref{VHS}) are shown in the figure (\ref{fig11}) as functions on $(v,w)$, for a value of $\gamma$ in the interval $(\gamma_{_{H}},-\gamma_{_{H}})$; similarly the polynomials (\ref{polymass}) are shown in the figure (\ref{polmass}).
The values $\pm \gamma_{_H}$ are critical, since the expressions (\ref{zerostable1}), and (\ref{zerostable2}) vanish, and thus the character of a local maximum for the zero-energy point, and the form of the potentials with stable minima  shown in the figure (\ref{fig11}) are not guaranteed;
therefore one must be careful by taking the limit $\gamma\rightarrow \pm \gamma_{_H}$.
Furthermore, the vacuum energies are given by
\begin{eqnarray}
V_R(\pm v_0, 0)= \frac{-3m^4_R}{2\lambda}\frac{P^v_R(\gamma)}{\gamma^2+2\gamma-1},\qquad V_H(\pm v_0, 0)= \frac{-3m^4_R}{2\lambda}\frac{2(\gamma^2-1)^2}{(\gamma^2+2\gamma-1)^2};
\label{vacenergy}
\end{eqnarray}hence, the depth of the red regions in the figure (\ref{fig11}) depends on $\gamma$; the polynomials  that deform the conventional vacuum energies in the above expressions are shown in the figure (\ref{vacenergy}). For $\gamma=0$ one can recover 
from $V_{R}$ the vacuum energy for the usual $U(1)$ field theory, since  $\frac{P^{v}_{R}(\gamma)}{(\gamma^2+2\gamma-1)}\Big|_{\gamma=0}=1$.
 In the sub-interval $[\gamma_{H},0]$ the deformation is small respect to the deformation in the sub-interval $[0,-\gamma_{H}]$; the 
 polynomial is varying continuously between the values $[.828,1]$ in the first sub-interval, and between the values $[1,0]$ in the second sub-interval, since $P^{v}_{R}(-\gamma_H)=0$.
 Furthermore, from the figure (\ref{vacenergy}) it is evident that the deformation of $V_H$ is strictly bigger than that of $V_{R}$; thus, the depth of the red regions in the figure (\ref{fig11}) is higher for $V_H$. Such a difference in the deformation is maximum for $-\gamma_{H}$, and minimum for 
 $\gamma_{H}$.
 \begin{figure}[H]
  \begin{center}
  \includegraphics[width=.6\textwidth]{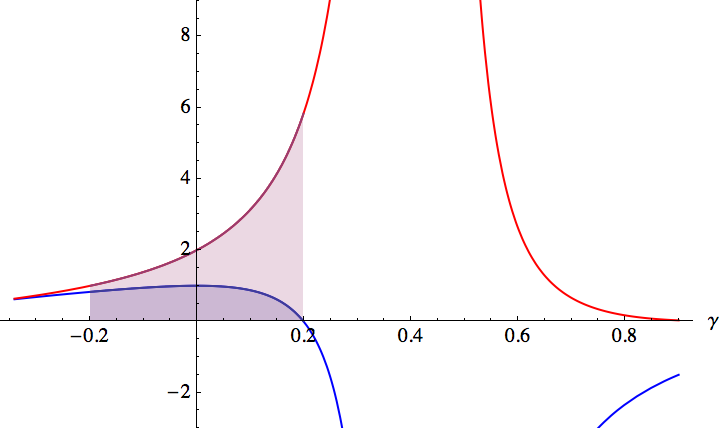}
\end{center}
\caption{The blue curve represents  $P^{v}_{R}(\gamma)/(\gamma^2+2\gamma-1)$; the red curve represents $\frac{2(\gamma^2-1)^2}{(\gamma^2+2\gamma-1)^2}$. In the interval $(\gamma_{_{H}},-\gamma_{_{H}})$ the polynomials take positive values, and the vacuum energies are finite, even at the limits $\pm \gamma_{_H}$.}
\label{vacenergy}
\end{figure}
The circular and hyperbolic rotations can be spontaneously broken by the choice
\begin{equation}
\sin(\theta-\theta_0)=0;\quad \sinh(\chi-\chi_0)=0; \label{43}
\end{equation}
in addition to  the vacuum expectation values  $(v_0\neq 0, w_0=0)$; thus all mixed terms $vw$ (both real and hybrid) can be gauged away, reducing the quadratic terms in Eq.\ (\ref{neve}) to the canonical form,
\begin{eqnarray}
      -\frac{am^2}{2}\psi\overline{\psi} + \frac{\lambda}{4!}(\overline{\psi}^{2}_{0} \psi^{2} + \psi^{2}_{0} \overline{\psi}^{2} ) &= &a\Big(\frac{1-\gamma^2}{2}m^2_R-\gamma m^2_H\Big)v^2 +a  \Big(\frac{1-\gamma^2}{2}m^2_R+\gamma m^2_H\Big)w^2 \nonumber\\&+&ija \Big[
 \left(\frac{1-\gamma^2}{2}m^2_H+\gamma m^2_R\right)v^2 + \left(\frac{1-\gamma^2}{2}m^2_H-\gamma m^2_R\right)w^2\Big]\nonumber\\
&=&ij\frac{a}{2}(-2P^v_H m^2_R)v^2;
\label{ssb4} 
\end{eqnarray} 
where the last equality follows from the mass relation (\ref{positivequan}); this expression must be compared with the mass terms in Eq.\ (\ref{VRS}), and (\ref{VHS}). Therefore, the field $w$ is massless in both senses, real and hybrid. The field $v$ has duplicated its hybrid mass with a change of sign; note that the real mass of $v$ has disappeared. The duplication of the mass with a change of sign for any $\gamma$ in the allowed interval is shown in the figure (\ref{polmassafter}).  
The mass that arises from SSB in Eq. (\ref{ssb4}), is actually a {\it running} mass, since the polynomial $P^v_H$ takes values in the interval $(0,-2.7165)$; hence the mass is running in the interval $(0, 2.7165m^2_R)$, from a nearly massless field to a ``heavy" field. Strict masslessness
is prohibited, since the value $-\gamma_{_H}$ is critical in relation to the form of the potentials with  local stable minima required for the spontaneous symmetry breaking.
\begin{figure}[H]
  \begin{center}
  \includegraphics[width=.6\textwidth]{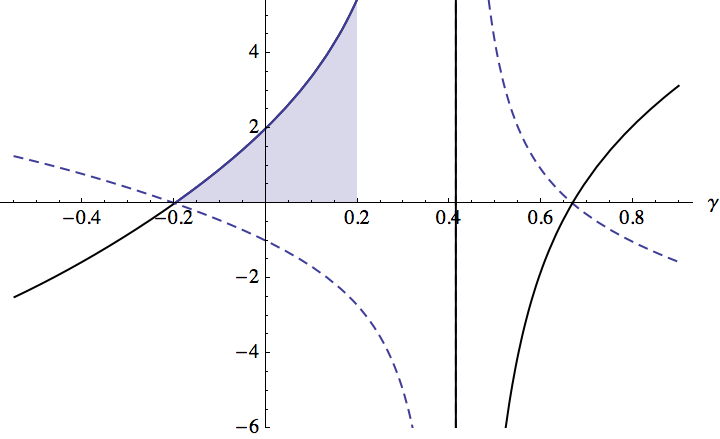}
\end{center}
\caption{ The continuous curve represents the mass polynomial coefficient $-2P^v_{H}(\gamma)$ after SSB: the dashed curve represents the mass polynomial coefficient $P^{v}_{H}(\gamma)$ before the SSB; the case $\gamma=0$ reproduces the usual SSB of $U(1)$. The field $w$ is a fully massless field for any $\gamma$ in the interval.}
\label{polmassafter}
\end{figure}
From the expression (\ref{thenorm}) we can realize that in the limit $\gamma\rightarrow 0$, the usual norm of an ordinary $U(1)$ complex field can be recuperated; in this sense the case with $\gamma\neq 0$ can be understood as a hyperbolic deformation around the usual formulation. Furthermore, according to the Eq. (\ref{positivequan}), in such a limit there is not difference between $m_R$ and $m_H$; the vaccum expectation value $v^2_0$ will reduce to the usual $U(1)$ expression. Similarly the inequality (\ref{ineq4}) will reduce to the inequality (\ref{ineq44}), which corresponds to the usual constraint in the $U(1)$ field theory. Likewise, all polynomials (\ref{polymass}) reduce to $-1$, and consequently there is no difference between $V_R$ and $V_H$. Therefore, the usual Mexican hat potential can be recuperated from the deformed version described in the figure (\ref{fig11}); this process is shown in the figure 
(\ref{nogamma1}).
\begin{figure}[H]
  \begin{center}
  \includegraphics[width=.5\textwidth]{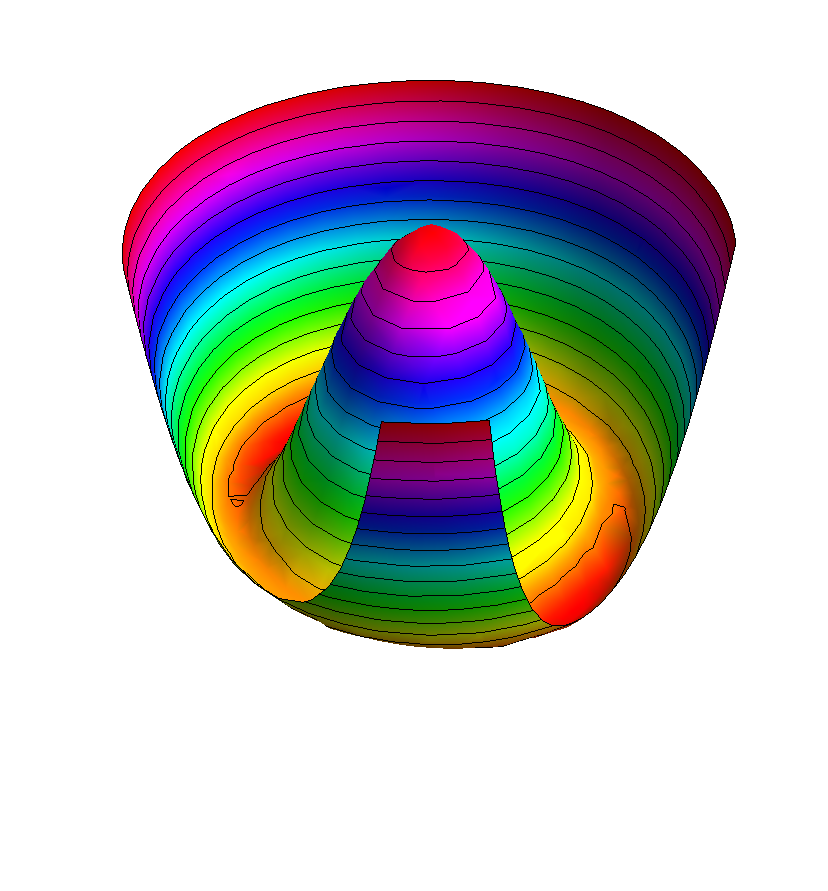}
\end{center}
\caption{The mexican hat potential as the limit of $V_{R}$ and $V_{H}$ as $\gamma\rightarrow 0.$}
\label{nogamma1}
\end{figure}

\subsection{The case $\gamma^2 \neq 1$, $(v_0= 0,w_0\neq 0)$, and $\lambda_{R}=\lambda_{H}$.}
\label{nogamma11}
\begin{figure}[H]
  \begin{center}
  \includegraphics[width=.6\textwidth]{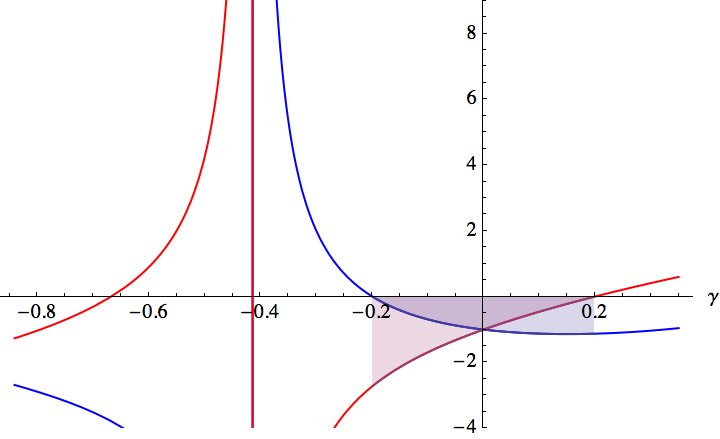}
\end{center}
\caption{$Q^w_R$ corresponds to the blue curve, with values in the interval $(0, -1.1252)$; $Q^w_H$ is represented in red, with values in the interval $(0, -2.7165).$}
\label{ssb44}
\end{figure}

\begin{figure}[H]
  \begin{center}
  \includegraphics[width=.6\textwidth]{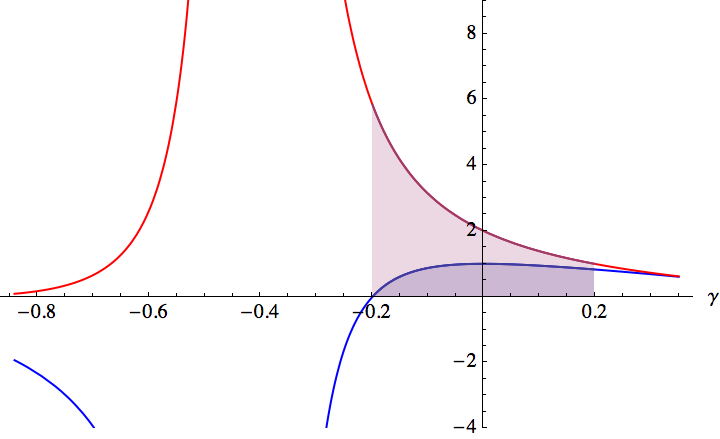}
\end{center}
\caption{The description of the vacuum energies given in the figure (\ref{vacenergy}) is essentially valid for this case, in relation to the behavior in the subintervals $[\gamma_H,0]$, and $[0,-\gamma_H]$, and the differences between the values of $V_R$ and $V_H$ in the vacuum.}
\label{vacenergy22}
\end{figure}

Along the same lines, the use of the Eqs. (\ref{masscocient2}), and (\ref{expvalue2}) lead
 essentially to the same expressions (\ref{positivequan}), (\ref{ineq4}), (\ref{zerostable1}), (\ref{zerostable2}), with the change $\gamma\rightarrow-\gamma$; the figure (\ref{poly4}) is essentially the same; for example the red and black curves will be convert to each other, without changing the interval $(\gamma_H, -\gamma_H)$; within this interval all new polynomials are negative. Furthermore, the inequality (\ref{ineq44}) remains valid, and the potentials can be described in terms of the same mass polynomials appearing in the expressions (\ref{VRS}) and (\ref{VHS});
\begin{eqnarray}
V_{R}= \frac{a}{2}Q^{v}_{R}m^2_{R}v^2+\frac{a}{2}Q^{w}_{R}m^2_{R}w^2+ \frac{\lambda}{6} \Big[ \cdots \Big]; 
     \label{VRSS} \\
V_{H}= \frac{a}{2}Q^{v}_{H}m^2_{R}v^2+\frac{a}{2}Q^{w}_{H}m^2_{R}w^2+ \frac{\lambda}{6} \Big[ \cdots \Big]; 
     \label{VHSS}    
     \end{eqnarray} 
 where the dots represent exactly the same expressions for the $\lambda$-terms in Eqs.\ (\ref{VRS}), and (\ref{VHS}); the polynomials $Q$ are defined as
 \begin{eqnarray}
 Q^{v}_{R}=P^w_R(\gamma\rightarrow-\gamma), \quad Q^{w}_{R}= P^v_R(\gamma\rightarrow-\gamma) ; \quad
  Q^{v}_{H}=Q^v_R, \qquad Q^{w}_{H}=  P^{v}_{H}(\gamma\rightarrow-\gamma); \label{polynomass}\end{eqnarray}
these potentials have the same form shown in the figure (\ref{fig11}).
However, in spite of the similarities, there will be an important difference; although the field $v$ will be massless as expected, the field $w$ will develop a mass with both parts, real and hybrid; explicitly we have that,
\begin{eqnarray}
      -\frac{am^2}{2}\psi\overline{\psi} + \frac{\lambda}{4!}(\overline{\psi}^{2}_{0} \psi^{2} + \psi^{2}_{0} \overline{\psi}^{2} ) = \frac{a}{2}m^2_R(-2Q^w_R)w^2+\frac{a}{2}ijm^2_R(-2Q^w_H)w^2;\label{double}
\end{eqnarray}
which are shown in the figure (\ref{ssb44}); the real mass in Eq. (\ref{double}) is running in the interval $(0, 1.1252m^2_R)$, and the hybrid mass
is running in the interval $(0, 2.7165m^2_R)$. Note that when the real mass goes to zero as $\gamma\rightarrow\gamma_H$, the hybrid mass goes to its maximum value, and in reverse in the limit $\gamma\rightarrow -\gamma_H$; the masses coincide in the value $m^2_R$ for $\gamma=0$.
In this case the vacuum energies are given by
\begin{eqnarray}
V_R(0,\pm w_0)= \frac{-3m^4_R}{2\lambda}\frac{P^v_R(-\gamma)}{\gamma^2-2\gamma-1},\qquad V_H(0,\pm w_0)= \frac{-3m^4_R}{2\lambda}\frac{2(\gamma^2-1)^2}{(\gamma^2-2\gamma-1)^2};
\label{vacenergy2}
\end{eqnarray}which can be obtained directly from the expression
(\ref{vacenergy}) through the change $\gamma\rightarrow-\gamma$; the polynomials in the expressions (\ref{vacenergy2}) are shown in the figure
(\ref{vacenergy22}).

 \subsection{The vacuum manifold for $\gamma\in (-\gamma_H, \gamma_H)$.}
\label{vm}
In the usual formulation of the theory with a conventional complex field, the degenerate vacuum is identified with the bottom of the mexican hat potential, the $S^1$ compact potential defined by the constraint $|v_0+i w_0|^2=\frac{6m^2}{\lambda}$, and parametrized by $v_0=\sqrt{\frac{6m^2}{\lambda}}\cos\theta$, and  $w_0=\sqrt{\frac{6m^2}{\lambda}}\sin\theta$; this case corresponds essentially to the choice $\gamma=0$ in the formulation at hand.  However, in the present case the hypercomplex field is undetermined additionally by a hyperbolic phase, leading to a two dimensional manifold for the degenerate vacuum; this case corresponds to the choice $\gamma\neq 0$ in the expression (\ref{correct}). We shall see that the vacuum manifold will correspond to a non compact space containing the hyperbola and the circle as factor spaces, and embedded in a four dimensional ambient space.

\begin{figure}[H]
  \begin{center}
    \includegraphics[width=.45\textwidth]{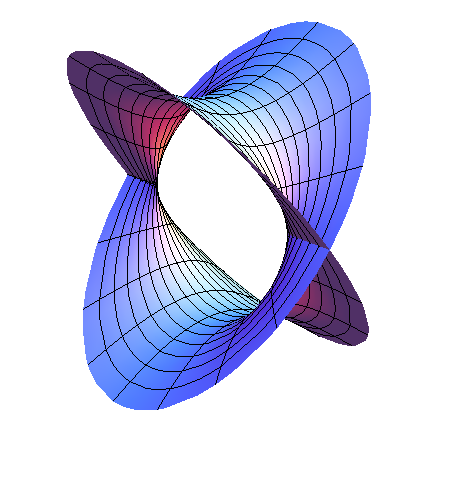}
  \caption{The degenerate vacuum as  a non-compact and not simply connected two manifold embedded in the 3d space. Usually only the compact transversal sections related with $U(1)$ have been considered as the vacuum; for this manifold of genus 1, we have that $\pi_0=1, \pi_1={\cal Z}$, and $\pi_{n\geqslant 2}=0$, 
 since it is homotopic to $S^1$. }  
   \label{klein}
  \end{center}
\end{figure}

A parametrization for the vacuum manifold can be given by the expression (\ref{combined}), with $\gamma\in (-\gamma_H, \gamma_H)$, and with  $\lambda \psi_{0}\overline{\psi}_{0}=-6am^2$; one can consider additionally the choice (\ref{expvalue}) with $w_{0}=0$,  and $v_{0}\neq 0$;
\begin{eqnarray}
\psi_{0}=v_0\Big\{(\cos\theta_{0}\sinh\chi_0-\gamma\sin\theta_{0}\cosh\chi_0)+i(\gamma\cos\theta_{0}\cosh\chi_0+\sin\theta_{0}\sinh\chi_0) \nonumber \\ +j(\cos\theta_{0}\cosh\chi_0-\gamma\sin\theta_{0}\sinh\chi_0)+ij(\sin\theta_{0}\cosh\chi_0+\gamma\cos\theta_{0}\sinh\chi_0)\Big\},
\label{cirhip}
\end{eqnarray}
 One can embed the vacuum manifold as a 2-dimensional subspace of the real four dimensional space defined by the four components of the field $\psi_{0}=x_0+iy_0+jz_0+ijw_0\rightarrow (x_0, y_0, z_0,w_0)$, and the result must be projected from 4D to 3D  in order to be visualized, and to gain insight about its geometrical and topological properties. The more practical and direct method that can be used is the projection of the 2-manifold into the 3-dimensional hyperplanes that define the four coordinate hyperplanes in the 4 dimensional ambient space, through $ (x_0, y_0, z_0,w_0)\rightarrow (x_0, y_0, z_0)$, and similarly into the other three hyperplanes.
Although in general such projections can be different depending of the {\it orientation} of the 2-manifold with respect to the coordinate hyperplanes, in this case, the four projections coincide to each other, and is represented in the figure \ref{klein}.
This projection is a sort of {\it product} of a hyperbola and a circle; it can be visualized also as two-dimensional planes embedded in 3-dimensions, and sharing a hole. The {\it self-intersection} is an effect of the projection into a 3-dimensional hyperplane; in the original four dimensional ambient space the two-manifold has no such self-intersections. This case is similar to the very known Klein bottle, which can not be realized in $R^3$ without intersecting itself.

We remark that the constraint defining the degenerate vacuum $\lambda \psi_{0}\overline{\psi}_{0}=-6am^2$, retains the full symmetry $SO^+(1,1)\times U(1)$, and the vacuum manifold is transformed into itself by the action of these transformations, and we have an infinite number of possible vacuum states with the same energy. The vacuum manifold is homotopic to  $S^1$, and the {\it string defects} can be form as topological defects; we shall analyze in detail this topic in sections  \ref{topdef} and
\ref{tdlocal}.

\subsection{ Polar parametrization for the fields}
\label{polar}

In the previous cases the hypercomplex field $\psi$ is described in terms of Cartesian components $(v,w)$ as real variable fields; now we use a {\it polar} decomposition, considering that we have at hand only two real variable fields. Thus, following the ideas at the end of the section \ref{hyrot}, for an expression of the form $(\rho_{_R}+ij\rho_{_H}) e^{i\xi} e^{j\eta}$, where $\rho_{_R}$, $\rho_{_H}$, $\xi$, and $\eta$, are real field variables, one must to choice two of them as  constants; similarly, the polar form for the vacuum field reads $\varphi_{0} = (\rho^R_{0}+ij \rho^H_{0}) e^{i\xi_{0}} e^{j\eta_{0}}$.
 
{\bf case $\gamma^2=1$}: For this case studied previously in the Section (\ref{gamma1}), we have that the norm of the dynamical field reduces to
$\psi\overline{\psi}=2ij\gamma(w^2-v^2)$; additionally the norm of the corresponding polar form will reduce to 
\begin{equation}
\psi\overline{\psi}=(\rho^2_{_R}-\rho^2_{_H})+2ij \rho_{_R}\rho_{_H};
\label{polarnorm}
\end{equation}
thus we have that $\rho_{_R}=\gamma\rho_{_H}$, and consequently 
\begin{eqnarray}
\psi= (\gamma+ij)\rho e^{i\xi} e^{j\eta}, \qquad \rho\equiv \rho_{_H}=\sqrt{w^2-v^2};
\label{polarform}
\end{eqnarray}
therefore, in this case only one degree of freedom may is encoded in the polar part, and the remaining degree of freedom will be encoded in one of the phases; similarly the vacuum field takes the form $\psi_0= (\gamma+ij)\rho_0 e^{i\xi_0} e^{j\eta_0}$, with the constraint $\rho^{2}_{0} = \frac{6m^{2}_{H}}{\lambda}$.

 We consider first the case $\psi (x) = (\gamma+ij)\rho (x) e^{i\xi (x)/\rho_0} e^{j\eta}$,  with $\eta$ constant, and excitations about the ground state with vanishing phases, $\xi_{0}=0=\eta_{0}$, and hence the circular and hyperbolic rotations are broken. Now, we write the dynamical field as $\big(\gamma+ij)(\rho (x) + \rho_{0}\big) e^{i\xi (x)} e^{j\eta}$, and the Lagrangian becomes
\begin{equation}
     {\cal L} = 2\gamma ij \int d x^{d} \Big [ \frac{1}{2} \partial^{i}\rho \partial_{i}\rho + \frac{1}{2}  \partial^{i}\xi \partial_{i}\xi - m_{_H}^{2} \rho^{2} + {\rm higher \ terms}\Big ] ,
     \label{rhomass}
\end{equation}
and, as expected, we have massive radial oscillations $\rho$ and circular {\it angle} massless oscillations $\xi$. There is not hyperbolic angle oscillations, due to the fixing $\eta = constant$, which has disappeared completely from the Lagrangian.
Similarly, if we fix $\xi = constant$, and $\eta \rightarrow \eta (x)/\rho_0$, then the oscillations around the same ground state are described by a Lagrangian of the form $L = \frac{1}{2} (\partial\rho)^{2} - \frac{1}{2}(\partial\eta)^{2} - m^{2}_H \rho^{2}$, plus higher terms, where the massive radial oscillations remain unchanged, but the hyperbolic angle oscillations appear now as massless modes with a global change of sign in the kinetic term. In these two cases, the massive radial oscillations are orthogonal, in field space, to the vacuum manifold described in the figure (\ref{klein}); furthermore, the angular oscillations are realized along of the direcctions defined by the decomposition of the vacuum manifold as product of a circle and a hyperbola.

 Now one can enforce the restriction $\rho= constant$, and thus the two degrees of freedom will be encoded in the phases, and all fluctuations will lie in the {\it valley} directions; such directions are defined by the circular and hyperbolic lines on the vacuum manifold, and the fluctuations will correspond to purely massless modes, as expected.  in this manner, If the two dynamical degrees of freedom are encoded in the phases, then
\begin{eqnarray}
\psi=(\gamma+ij)\rho_{0}e^{i\xi(x)/\rho_0}e^{j\eta(x)/\rho_{0}} \!\! & = & \!\! (\gamma+ij) \rho_0\Big(1+i\frac{\xi(x)}{\rho_0}+\cdots\Big)\Big(1+j\frac{\eta(x)}{\rho_{0}}+\cdots\Big),\nonumber \\
 \!\! & = & \!\!(\gamma+ij)[{\rho_0}+i\xi(x)+j\eta(x)+\cdots]
  \label{valley}
\end{eqnarray}
where $\xi$, $\eta$, as well as their derivatives, are considered
to be small real fields, and the dots represent higher order terms in the perturbations; therefore, the substitution into the Lagrangian yields the expression,
\begin{equation}
{\cal L}(\xi,\eta) = \gamma ij \int d x^{d} \Big [\partial^{i}\xi \partial_{i}\xi -\partial^{i}\eta \partial_{i}\eta-2ij\partial^i\xi\partial_i\eta + {\rm higher \ terms}\Big ],
     \label{valley1}
\end{equation}
since the mass terms have disappeared completely, the fields excitations in the {\it valley} directions are massless modes. Note however that the kinetic terms have a nonconventional form, due to the presence of a hybrid term that mixes the gradients of the field excitations $(\xi,\eta)$; however, one can interchange between the canonical form of quadratic gradients and the mixed form through the invertible mapping
\begin{eqnarray}
     \left( \begin{array}{c} \xi\\ \eta \end{array} \right)\leftrightarrow \frac{1}{2}\left( \begin{array}{cc}
                 1 & -1 \\
                 1 & 1
                 \end{array} \right)\left( \begin{array}{c} \xi\\ \eta \end{array} \right),  \quad \partial^i\xi\partial_i\eta \leftrightarrow \partial^{i}\eta \partial_{i}\eta-\partial^{i}\xi \partial_{i}\xi.
     \label{h}
\end{eqnarray}
Alternatively, the kinetic terms in the Lagrangian (\ref{valley1}) can be rewritten in terms of the derivatives of the full phase $i\xi+j\eta$, by considering the identity $[\partial^{k}(i\xi+j\eta)][\partial_{k}(i\xi+j\eta)]=-[\partial^{k}(i\xi+j\eta)]\overline{[\partial_{k}(i\xi+j\eta)]}=-[\partial^{i}\xi \partial_{i}\xi -\partial^{i}\eta \partial_{i}\eta-2ij\partial^i\xi\partial_i\eta]$, which corresponds essentially to the kinetic terms in Eq.(\ref{h}).
Furthermore, since we are considering small oscillations around a point of the vacuum manifold, the dynamics of the circular and hyperbolic oscillations are locally indistinguishable; in fact the expressions (\ref{valley}), (\ref{valley1}), and (\ref{h}) are invariant under the interchange $i\xi\leftrightarrow j\eta$. As an option, one can distribute the global hybrid term $ij$ out of the integration in the Lagrangian (\ref{valley1}), and hence the real and hybrid terms under integration will interchange their roles, without changing the physical conclusions. 

{\bf case $\gamma^2\neq 1$}: In this case the comparison between the norm $\psi\overline{\psi}=(\gamma^2-1)(v^2+w^2)+2ij\gamma(w^2-v^2)$, and the expression (\ref{polarnorm}) leads to  
\begin{eqnarray}
\rho^2_{_R}-\rho^2_{_H}=(\gamma^2-1)(v^2+w^2), \quad  \rho_{_R}\rho_{_H}=\gamma (w^2-v^2);
\label{doublepolar}
\end{eqnarray}
and similarly for the vacuum fields; these maps allow us to describe the fields with the pair $( \rho_{_R},\rho_{_H})$ instead of the pair $(v,w)$; 
now, we rewrite the dynamical field as $[ \rho_{_R}+\rho^{R}_{0}+ij(\rho_{_H}+\rho^{H}_0)]$, leading to a Lagrangian of the form
\begin{equation}
{\cal L}( \rho_{_R},\rho_{_H}) = \frac{1}{2}\int d x^{d} \Big [\partial^{i}\rho_{_R} \partial_{i}\rho_{_R} -\partial^{i}\rho_{_H} \partial_{i}\rho_{_H}+2ij\partial^i\rho_{_R}\partial_i\rho_{_H} +2m^2(\rho^2_{_R}-\rho^2_{_H}+2ij\rho_{_R}\rho_{_H}) +{\rm higher \ terms}\Big ];
     \label{radialmassive}
\end{equation}
with $m^2=m^2_R+ijm^2_H$; hence, the two radial modes are massive as expected. This expression can be rewritten in a compact form in terms of the Hermitian field $\Pi\equiv \rho_{_R}+ij\rho_{_H}$, 
\begin{equation}
{\cal L}( \rho_{_R}+ij\rho_{_H}) = \frac{1}{2}\int d x^{d} \Big [\partial^{i}\Pi \partial_{i}\Pi  +2m^2\Pi^2+{\rm higher \ terms}\Big ];
     \label{radialmassive2}
\end{equation}
therefore, this Hermitian field $\Pi$ encode two massive radial oscillations that are orthogonal to the two-dimensional vacuum manifold; the corresponding mass is also Hermitian.

All these scalar fields,  massless and massive bosons will be completely eaten by a vector field through Higgs mechanism, once we consider the coupling to hypercomplex  QED and local rotations; this will lead to massive pure electrodynamics, without the presence of Higgs massive fields.

\subsection{ Formation of global topological strings: confirming the Derrick's theorem}
\label{topdef}
Field configurations that define topological defects correspond to domains where the symmetry is left unbroken, {\it i.e.} satisfy the constraint $\phi=0$; we consider for example the case of a four dimensional space-time $(x,y,z,t)$ as background; for time independent field configurations of the form (\ref{ring}), $\phi=\phi_1+i\phi_2+j\phi_3+ij\phi_4$, such a constraint implies that $\phi_1(x,y,z)=\phi_2(x,y,z)=\phi_3(x,y,z)=\phi_4(x,y,z)=0$, which define  monopoles as  possible topological defects. In the same background a conventional complex field of the form 
$\phi=\phi_1+i\phi_2$, leads to string defects. Therefore, a hypercomplex field of the form (\ref{ring}) leads to monopoles in a five-dimensional background, and to string defects in a six-dimensional background. However, for a hypercomplex field of the form (\ref{correct}) defined in terms of two real functions, we have again string defects as possible topological defects in a four dimensional space-time, such as a conventional complex field. These algebraic constraints complemented with a vacuum manifold with nontrivial fundamental group will yield the possibility of formation of string defects in the theories considered in this paper.

Let us remember, following \cite{kibble,olesen,vilenkin}, the relevant aspects for the case at hand, of the usual description of the formation of string defects or vortices in three spatial dimensions in a theory with a U(1) global symmetry; the minima lie on a circle with $\psi_0= \sqrt{\frac{6m^2}{\lambda}}e^{i\alpha}$, and around a loop in the three-space, the phase $\alpha$ takes values in $[0,2\pi]$, and by continuity, within the loop the field must vanish $\psi=0$; the locus of such points lies on the core of the string defect, constrained to be one-dimensional in three spatial dimensions. Now, the condition that the field must be zero within a loop, can be associated with some value of the hyperbolic parameter for the case at hand, say $\chi=0$; thus, one can relate each point in the locus to each value of $\chi$ in the intervale $(-\infty,+\infty)$, leaving intact the formation of an infinite defect of false vacuum points. Therefore, by enlarging the usual $U(1)$ vacuum manifold to a cylindrical manifold, the string defect will lie over the new non-compact transversal direction; the formation of this defect is consistent with the fact that the cylindrical manifold is homotopic to $S^1$; hence, topologically distinct string defects are labelled by the same elements of the fundamental group of $S^1$, the usual winding number $n\in {\cal Z}-\{0\}$. Now, in the usual description with a $U(1)$ global symmetry,  a string defect with the core aligned with the $z$-axis, in cylindrical polar coordinates $(r,z,\theta)$, the field asymptotically takes the form 
\begin{equation}
\lim_{r\rightarrow+\infty}\psi(r,z,\theta)=\rho_{0}e^{in\theta},
\quad \rho_0= \sqrt{\frac{6m^2}{\lambda}},
\label{ti}
\end{equation} 
and for the case at hand we have the extended version that incorporates a hyperbolic phase that breaks the translational invariance in the noncompact $z$-direction of the expression (\ref{ti}),
\begin{equation}
\lim_{r\rightarrow+\infty}\psi(r,z,\theta)=\rho_{0}e^{in\theta}e^{jlz}, \quad \rho_{0}=\rho^R_{0}+ij \rho^H_{0};
\label{limit}
\end{equation}where $lz$ is adimensional in natural units; with this condition we identify the asymptotic form of the field with its ground state.

Now, following the usual treatment for string defects, we look for a static exact solution for the equations of motion (\ref{em}) with the ansatz 
\begin{equation}
     \psi = \rho_{0} f(\rho_0 r) e^{in\theta} e^{jlz}, \qquad f(0)=0, \qquad \lim_{r\rightarrow +\infty} f(\rho_0r) = 1;
     \label{static}
\end{equation}
where $\rho_0 r$, and $lz$ are adimensional variables in natural units; the asymptotic limit is required by consistency with the condition (\ref{limit}). With the substitution into the equations of motion (\ref{em}), these reduce to a non-linear ordinary equation for $f$;
\begin{equation}
     f'' + \frac{1}{\rho_0 r} f' - \frac{n^{2}}{(\rho_0 r)^{2}} f - (f^{2}-1) f + \frac{l^2}{\rho^2_0}f =0,     
     \label{f}
\end{equation}
with the exception of the last term $f$ that comes from the new term $f \partial^{2}z e^{jlz}$, all terms correspond to the usual ones in the traditional scheme; hence, the approximate asymptotic solutions remain essentially the same,
\begin{eqnarray}
     f(\rho_0 r) \!\! & \approx & \!\! C_{n} r^{n} + \cdots , \qquad r\rightarrow 0, \nonumber \\
     f(\rho_0 r) \!\! & \approx & \!\! 1- {\cal O}(r^{-2}), \qquad r\rightarrow\infty ,
     \label{flimit}
\end{eqnarray}
Now the gradient of the field is
\begin{equation}
     \nabla\psi = \rho_{0} \Big[f' \widehat{r} + in\frac{f}{r}\widehat{\theta} + jlf\widehat{z}\Big] e^{in\theta} e^{jlz},
     \label{gradientfield}
\end{equation}
with a new contribution in the $\widehat{z}$-direction; thus the gradient energy density is
\begin{equation}
     |\nabla\psi|^{2} = \rho^{2}_{0} \Big[(f')^{2} + n^{2} \frac{f^{2}}{r^{2}} - l^{2}f^{2}\Big];
     \label{gradientenergy}
\end{equation}
hence, the energy per unit length is given by
\begin{equation}
     2\pi\rho^{2}_{0} \Big(n^{2} \int^{\infty} \frac{dr}{r} - l^{2} \int^{\infty} r dr\Big),
     \label{derrick}
\end{equation} 
where we have the usual logarithmical contribution to an infinite-energy, and additionally we have a new contribution with a quadratic divergence, which is worse than the logarithmical contribution. Therefore, the energy is infinite, in fact tending to $-\infty$ instead to $+\infty$ as in the usual case; this result is consistent with the Derrick's theorem \cite{derrick}, which establishes that
there are not finite-energy, time independent solutions, with scalar fields only, that are localized in more than one dimension. However,  in the present formulation such a  divergence can be cured by considering compensating gauge fields (see Section \ref{tdlocal}), such as in the usual $U(1)$ global strings.

\section { Hypercomplex electrodynamics: local symmetries}  
\label{hed}

For the usual formulation that describes a charged scalar field couple to QED, we have the Lagrangian
\begin{equation}
     {\cal L} = - \frac{1}{4} F^{2}_{\mu\nu} +  | (\partial_{\mu} - ieA_{\mu})\psi|^{2}-V(\psi,\overline{\psi}),
     \label{QED}
\end{equation}
with two coupling constants $e$ and $\lambda$;
the hyperbolic rotations can be incorporated as a part of the local gauge symmetry of the Lagrangian (\ref{QED}), by considering the expression (\ref{correct}) for the hyper-complex extension of the modulus $\psi\cdot\overline{\psi}$, and the local gauge transformations
\begin{equation}
     \psi \rightarrow e^{i\theta} e^{j\chi} \psi, \qquad A_{\mu} \rightarrow A_{\mu} + \frac{1}{e} (\partial_{\mu}\theta - ij\partial_{\mu}\chi), \quad F_{\mu\nu}= \partial_{\mu}A_{\nu}-\partial_{\nu}A_{\mu} \rightarrow F_{\mu\nu},     
               \label{hgt}
\end{equation}
where in general the arbitrary real functions depend on the background space-time coordinates, $\theta = \theta (x)$, $\chi =\chi (x)$. Furthermore, the Eqs. (\ref{hgt}) imply that $\overline{A}_{\mu} \rightarrow \overline{A}_{\mu} + \frac{1}{e} (\partial_{\mu}\theta - ij\partial_{\mu}\chi)$, and thus, the condition $A_{\mu} - \overline{A}_{\mu}=0$, is preserved under gauge transformations, in spite of the hyper-complex extension of the fields; hence the vector potential is ``Hermitian" in the hypercomplex sense.

The equations of motion and the energy-momentum tensor for the action (\ref{QED}) are,
\begin{eqnarray}
     \partial^{\mu} F_{\mu\nu}= ie (\psi\overline{D_{\nu}\psi} - \overline{\psi} D_{\nu}\psi), \label{eqem1}\\
     (\partial^{\mu} - ie A^{\mu})(\partial_{\mu} - ieA_{\mu})\psi + \frac{\lambda}{12} (\psi\overline{\psi} - \frac{6m^{2}}{\lambda})\psi = 0, \label{eqem2}\\
     T_{\mu\nu} =  D_{(\mu}\psi \cdot \overline{D_{\nu )}\psi} - \frac{1}{2} F_{\mu}{^{\alpha}} F_{\nu\alpha} - \frac{1}{2} g_{\mu\nu} {\cal L}.
     \label{emtqed}
\end{eqnarray}
Similarly in this case, the states with minimal energy are given by $\psi_{0} \overline{\psi}_{0} = \frac{6m^{2}}{\lambda}$, and the vector potential is pure gauge, $A^{0}_{\mu} = \frac{1}{e} (\partial_{\mu} \theta_{0} - ij\partial_{\mu}\chi_{0})$, with $\theta_{0}$ and $\chi_{0}$ arbitrary space-time dependent functions; hence, for vacuum fields we must have that $\nabla^{0}\psi^{0} \equiv \partial_{\mu} \psi^{0} - ieA^{0}_{\mu} \psi^{0}=0$, identically, which will be used implicitly below. Hence, the degenerate vacuum is essentially the manifold described in fig. (\ref{klein}).

First, we use the parametrization of the fields $\psi$ and $\psi_{0}$ given in Eq. (\ref{combined}); the expansions  (\ref{neve}), (\ref{quacub1}), and (\ref{quacub2}),  remain valid in the case at hand since do not contain the (covariant) derivatives of the fields; one only requires the switching on the space-time dependence of the phases $(\chi, \chi_{0}; \theta, \theta_{0})$.  Additionally, one must develop the expansion of the first two terms in Eq. (\ref{QED}); the expanding around the vacuum requires $\psi \rightarrow \psi + \psi_{0}$, and $A_{\mu} \rightarrow A_{\mu}+ A^0_{\mu}$, and thus the covariant derivative $(\partial_{\mu} -ieA_{\mu})\psi \rightarrow (\partial_{\mu} -ieA_{\mu})\psi -ie (A_{\mu}\psi_{0} + A^{0}_{\mu}\psi)$; the Lagrangian (\ref{QED}) reads
\begin{eqnarray}
     {\cal L} (\psi + \psi_{0}, A+A_{0}) \!\! & = & \!\! -\frac{1}{4} F^{2}_{\mu\nu} + e^{2}|\psi_{0}|^{2} B^{2}_{\mu} + \partial_{\mu}\psi \cdot \partial^{\mu} \overline{\psi} + B^{\mu} \Big \{ie(\overline{\psi}_{0}\partial_{\mu}\psi - \psi_{0}\partial_{\mu}\overline{\psi}) + 2e |\psi_{0}|^{2}(\partial_{\mu}\theta - ij \partial_{\mu}\chi) \nonumber \\
     \!\! & & \!\! + A^{0}_{\mu}(\psi_{0}\overline{\psi} + \overline{\psi}_{0}\psi)\Big\} +i (\partial^{\mu}\theta -ij\partial^{\mu}\chi)(\overline{\psi}_{0}\partial_{\mu}\psi - \psi_{0}\partial_{\mu}\overline{\psi})+ieA^{\mu}_{0} (\overline{\psi}\partial_{\mu}\psi - \psi\partial_{\mu}\overline{\psi})\nonumber \\
     \!\! & & \!\! + |\psi_{0}|^{2}(\partial_{\mu}\theta - ij\partial_{\mu}\chi) (\partial^{\mu}\theta -ij\partial^{\mu}\chi) -V(\psi+ \psi_{0},\overline{\psi}+\overline{\psi_0})+ {\rm higher \ terms},
     \label{qed2}
\end{eqnarray}
where $A_{\mu}$ has been replaced by $A_{\mu} = B_{\mu} + \frac{1}{e} (\partial_{\mu}\theta - ij\partial_{\mu}\chi)$, and $F_{\mu\nu}$ is expressed now in terms of the new field $B_{\mu}$. Additionally, we have that
\begin{eqnarray}
     \partial_{\mu}\psi \cdot \partial^{\mu}\overline{\psi} \!\! & \approx & \!\!  (\gamma^2-1)(\partial v^{2} + \partial w^{2} )+2ij\gamma (\partial w^{2} - \partial v^{2} ),    
     \nonumber \\
    \overline{\psi}_{0}\partial_{\mu}\psi - {\psi}_{0} \partial_{\mu}\overline{\psi} \!\! & \approx & \!\! 
    2i(\gamma^2+1)(w_{0}\partial_{\mu} v-v_0\partial_{u}w) ;
     \label{qed3} 
     \end{eqnarray}     
Relevant terms in the expression (\ref{neve}) that lead to quadratic terms in the Lagrangian (\ref{qed2}) are given, to second order, by the expressions (\ref{ssb4}), and (\ref{double}), by approaching the circular and hyperbolic phases to first order,
\begin{equation}
-\frac{am^2}{2}\psi\overline{\psi} + \frac{\lambda}{4!}(\overline{\psi}^{2}_{0} \psi^{2} + \psi^{2}_{0} \overline{\psi}^{2} ) \approx \left \lbrace
\begin{array}{ll}
-aijP^v_H m^2_Rv^2+ \cdots ; &  (v_{0}\neq 0, w_{0}=0) \\
-am^2_RQ^w_Rw^2-aijm^2_RQ^w_Hw^2 +\cdots;  & (v_{0}= 0, w_{0}\neq 0) \end{array}\right.;
\nonumber
     \label{qed4}
\end{equation}
now, the vanishing requirement of interaction terms of the form $B^{\mu} \cdot \partial_{\mu} (\varphi , \overline{\varphi}, \theta , \chi)$, in Eq. (\ref{qed2}) yields 
\begin{eqnarray}
\theta =\frac{\gamma^2+1}{|\psi_{0}|^{2}} (w_{0}v-v_{0}w),  \quad \chi=0;
\label{thetavw}
\end{eqnarray}
which corresponds to exploit the freedom of choosing the original gauge field $A_{\mu}$; the inverse expre\-ssion $\frac{1}{|\psi_{0}|^{2}}$ is very complicated, but fortunately we shall require only its values when only one of the scalar fields acquires a non-zero expectation value,
\begin{equation}
\frac{1}{|\psi_{0}|^{2}} = \left \lbrace
\begin{array}{ll}
\frac{\gamma^2-1+2ij\gamma}{(\gamma^2+1)^2v^2_{0}}; &  (v_{0}\neq 0, w_{0}=0) \\
\frac{\gamma^2-1-2ij\gamma}{(\gamma^2+1)^2w^2_{0}};  & (v_{0}= 0, w_{0}\neq 0) \end{array}\right.;
\nonumber
     \label{inversevacuum}
\end{equation}
thus, the quadratic terms in the Lagrangian reduce to
\begin{eqnarray}
     {\cal L} (\psi +\psi_{0}, A+A_{0}) \!\! & = & \!\! -\frac{1}{4} F^{2}_{\mu\nu} + e^{2}|\psi_{0}|^{2} B_{\mu}B^{\mu} + 
      (\gamma^2-1)(\partial v^{2} + \partial w^{2} )+2ij\gamma (\partial w^{2} - \partial v^{2} )\nonumber\\    
     \!\! & +& \!\!  \frac{(\gamma^2+1)^2}{|\psi_{0}|^{2}}\Big[w_{0}^2\partial v^{2}-v_0 w_{0}\partial^{\mu}v\partial_{\mu}w+v_{0}^2\partial w^{2}\Big] \nonumber \\
     \!\! & & \!\!-\Big[ -\frac{am^2}{2}\psi\overline{\psi} + \frac{\lambda}{4!}(\overline{\psi}^{2}_{0} \psi^{2} + \psi^{2}_{0} \overline{\psi}^{2} )\Big] + {\rm higher \ terms};    
     \label{exploitt}
\end{eqnarray}
in this expression, we have not yet fixed one of the vacuum fields $(v_{0},w_{0})$, and we have remanent mixed terms of the form $\partial^{\mu}v\cdot\partial_{\mu}w$, and $v\cdot w$, which turn out to be proportional to $v_{0}\cdot w_{0}$. Since only one v.e.v., $v_{0}$ or $w_{0}$ will acquire a non-zero value, then such mixed terms will vanish at the end; now we shall study each case separately.

 \subsection{The case ($v_{0}\neq 0, w_{0}=0)$}
 
\label{qedvacuum}  
\begin{figure}[H]
  \begin{center}
  \includegraphics[width=.6\textwidth]{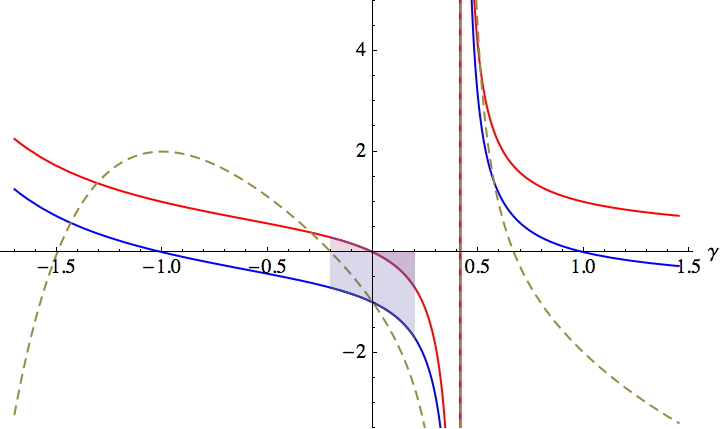}
\end{center}
\caption{ The blue curve represents $M^B_{R}(\gamma)$ , and the red curve $M^B_{H}(\gamma)$; the dashed curve represents $P^{v}_{H}(\gamma)$ in Eq. (\ref{exploit}); the case $\gamma=0$ reproduces the usual SSB of $U(1)$, with $P^{v}_{H}(0)=-1=M^B_{R}(0)$, and $M^B_{H}(0)=0$.}
\label{massBB}
\end{figure}

Using the expressions (\ref{qed4}), and (\ref{inversevacuum}), the Lagrangian (\ref{exploitt}) reduces to
\begin{eqnarray}
     {\cal L} (\psi +\psi_{0}, A+A_{0}) \!\! & = & \!\! -\frac{1}{4} F^{2}_{\mu\nu} + e^{2}|\psi_{0}|^{2} B_{\mu}B^{\mu} + 
      (\gamma^2-1-2ij\gamma)\partial v^{2}+aijP^v_H m^2_Rv^2 \nonumber \\  
       \!\! & + & \!\! {\rm higher \ terms};    
     \label{exploit}
\end{eqnarray}
the kinetic and the mass terms for the field $w$ have disappeared,  and corresponds thus to a Nambu-Goldstone field. Additionally we have a residual massive field $v$ with a non-zero vacuum expectation value  given by the expression (\ref{positivequan}); the v.e.v. of this Higgs field
 determines the (Hermitian) mass of the longitudinal mode of the (Hermitian) vector field $B$,
 \begin{eqnarray}
  e^{2}|\psi_{0}|^{2}=e^2(\gamma^2-1-2ij\gamma)v^2_{0}=\frac{6e^2}{a\lambda}\Big[ M^{B}_{R}(\gamma)+ijM^{B}_{H}(\gamma)
  \Big]m^2_{_{R}}; \nonumber\\
  M^{B}_{R}(\gamma)\equiv \frac{1-\gamma^2}{\gamma^2+2\gamma-1} , \quad M^{B}_{H}(\gamma)\equiv \frac{2\gamma}{\gamma^2+2\gamma-1} ;   \label{massesB}
   \end{eqnarray}
therefore, the Hermitian vector field $B$ that in general has the form $B\equiv B_{R}+ijB_{H}$, has acquired a Hermitian mass through Higgs mechanism; hence, one has two real masses for two real fields $(B_{R},B_{H})$.
The flows of the masses defined by these polynomials are shown in the figure \ref{massBB}; this figure shows the global behavior, and the behavior in the interval $(\gamma_{H},-\gamma_{H})$. The flows have the same asymptote, the root $\sqrt{2}-1$ of the polynomial $(\gamma^2+2\gamma-1)$. The real mass of $B$ vanishes at two roots of $M^{B}_{R}$,  $\gamma^2=1$, the purely hyperbolic limit for the theory; however,  these roots are out of the interval $(\gamma_{H},-\gamma_{H})$. 

The figure also shows that the Higgs boson field $v$ may have a mass as small as $\gamma\rightarrow \gamma_{H}$, although strict masslessness is prohibited; hence, in this limit one could expect a nearly massless Higgs boson. In this limit the vectorial field $B$ will acquire the smaller masses, 
\begin{eqnarray}
\frac{6e^2}{a\lambda}\Big[ M^{B}_{R}(\gamma_{H})+ijM^{B}_{H}(\gamma_{H})
  \Big]m^2_{_{R}}\approx \frac{6e^2}{a\lambda}\Big[ -0.7071+0.2929ij
  \Big]m^2_{_{R}}.
    \label{smaller}
    \end{eqnarray}
Similarly in the limit  $\gamma\rightarrow -\gamma_{H}$, the figure shows that the fields will acquire the higher masses,
\begin{eqnarray}
\frac{6e^2}{a\lambda}\Big[ M^{B}_{R}(-\gamma_{H})+ijM^{B}_{H}(-\gamma_{H})
  \Big]m^2_{_{R}}\approx \frac{6e^2}{a\lambda}\Big[ -1.7071-.7071ij
  \Big]m^2_{_{R}}.
    \label{higher}
    \end{eqnarray} 
The limit $\gamma\rightarrow \gamma_{H}$, is the limit of light masses for the fields, and  $\gamma\rightarrow -\gamma_{H}$ corresponds to the limit for massive fields; note that the difference between such limits is one mass unit for both, real and hybrid components.

\subsection{The case ($v_{0}= 0, w_{0}\neq 0)$}
\label{qedvacuum2} 
 
\begin{figure}[H]
  \begin{center}
  \includegraphics[width=.5\textwidth]{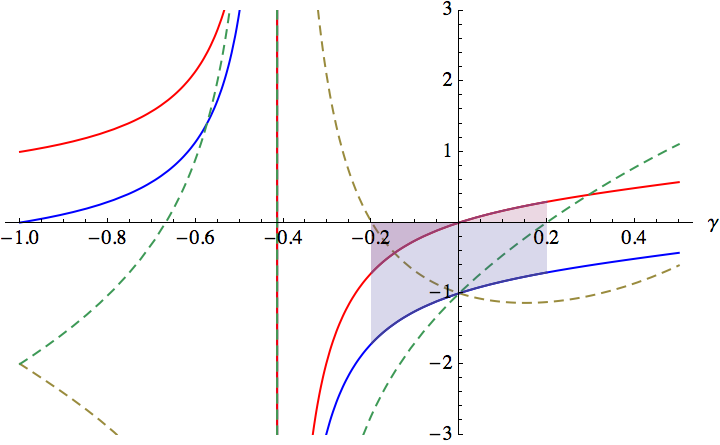}
\end{center}
\caption{ The mirrored image of figure \ref{massBB}.}
\label{massBB2}
\end{figure}

In this case, the Lagrangian (\ref{exploitt}) reduces to
\begin{eqnarray}
     {\cal L} (\psi +\psi_{0}, A+A_{0}) \!\! & = & \!\! -\frac{1}{4} F^{2}_{\mu\nu} + e^{2}|\psi_{0}|^{2} B_{\mu}B^{\mu} + 
      (\gamma^2-1+2ij\gamma)\partial w^{2}+am^2_RQ^w_Rw^2+aijm^2_RQ^w_Hw^2 ; \nonumber \\  
       \!\! & + & \!\! {\rm higher \ terms};    
     \label{exploit2}
\end{eqnarray}
now the kinetic and the mass terms for the field $v$ have disappeared,  and corresponds thus to a Nambu-Goldstone field. The residual field $w$ is massive in both sense, real and hybrid; the non-zero vacuum expectation value for this field is given by $w^2_{0}=\frac{6m^2_{_R}}{a\lambda(1-\gamma^2+2\gamma)}$ (see Section(\ref{nogamma11})), and determine the masses for the vectorial field $B$ shown in the figure (\ref{massBB2}).
This figure is basically the mirrored image respect to the $`y"$-axis of the figure \ref{massBB}; the qualitative and quantitative aspects of the expressions (\ref{smaller}), and (\ref{higher}) remain valid, and one only requires interchange the roles of the limits $\gamma_{H}\leftrightarrow -\gamma_{H}$. The difference respect to the Lagrangian (\ref{exploit}), is that the Higgs field $w$ appears with real and hybrid masses, and their flows are shown as dashed curves; this behavior was discussed in the figure \ref{ssb44}, and hence in the limit $\gamma\rightarrow \gamma_{H}$, the Higgs field $w$ will have a light real mass, and hybrid mass with its maximum value, and in reverse in the limit $\gamma\rightarrow -\gamma_{H}$.
 \subsection{ Polar parametrization for the fields: purely massive electrodynamics, no massive Higgs bosons}
\label{polar1}  
Along the lines followed in Sec. \ref{polar}, we consider the polar decomposition for the fields with radial and circular modes; 
 in the expression of the form $(\rho_{_R}+ij\rho_{_H}) e^{i\xi} e^{j\eta}$, for the dynamical field we can consider for generality that the four variables $\rho_{_R}$, $\rho_{_H}$, $\xi$, and $\eta$, are  space-time coordinates dependent, and at the end we shall consider that two of them must be constants. Hence, one has the following expressions;
\begin{eqnarray}
     \partial_{\mu}\psi\cdot\partial^{\mu} \overline{\psi} \!\! & = & \!\! (\partial\rho_{R}+ij\partial\rho_{H})^{2} -(\rho_{R}+ij\rho_{H})^{2} (i\partial\xi+i\partial \eta)^{2}, \nonumber \\
     \overline{\psi}\partial_{\mu}\psi - \psi\partial_{\mu}\overline{\psi} \!\! & = & \!\! 2(\rho_{R}+ij\rho_{H})^{2} (i\partial\xi+i\partial \eta), \nonumber \\
     V(\psi +\psi_{0}, \overline{\psi}+\overline{\psi}_{0}) \!\! & = & \!\! m^{2} \rho^{2} + {\rm higher \ terms},
     \label{approx}
\end{eqnarray}
and then
\begin{eqnarray}
 | (\partial_{\mu} - ieA_{\mu})\psi|^{2} \!\! & = & \!\! (\partial\rho_{R}+ij\partial\rho_{H})^{2} +e(\rho_{R}+ij\rho_{H})^{2}\Big[eB^{\mu}B_{\mu}+2B^\mu\partial_{\mu}[\theta-\xi+ij(\eta-\chi)] \nonumber\\
\!\! & -& \!\! \frac{1}{e}(i\partial\xi+i\partial \eta)^{2} +\frac{1}{e}\partial^{\mu}(\theta-ij\chi)\cdot\partial_{\mu}[\theta-2\xi+ij(2\eta-\chi)]\Big] , 
 \label{approx1}
 \end{eqnarray}
where $A_{\mu} = B_{\mu} + \frac{1}{e} (\partial_{\mu}\theta - ij\partial_{\mu}\chi)$; fluctuations around the vacuum require to rewrite the dynamical field as $\psi \rightarrow (\rho_{_R}+\rho^0_{_R}+ij(\rho_{_H}+\rho^0_{_H})) e^{i\xi} e^{j\eta}$, 
which inserted into the above expressions leads to,
\begin{eqnarray}
     {\cal L} (\psi +\psi_{0}, A+A_{0})\!\! & = & \!\! - \frac{1}{4} F^{2} + e^{2}  |\psi_{0}|^{2} B^{2}+(\partial\rho_{R}+ij\partial\rho_{H})^{2}- |\psi_{0}|^{2} (i\partial\xi+i\partial \eta)^{2}\nonumber\\  
    \!\! & + & \!\! 2e|\psi_{0}|^{2} B^\mu\partial_{\mu}[\theta-\xi
+ij(\eta-\chi)]\nonumber\\  
 \!\! & + & \!\!|\psi_{0}|^{2}\partial^{\mu}(\theta-ij\chi)\cdot\partial_{\mu}[\theta-2\xi+ij(2\eta-\chi)]\nonumber \\        \!\! & + & \!\!  am^{2}(\rho_{R}+ij\rho_{H})^{2} + {\rm higher \ terms}.
     \label{local1}
\end{eqnarray}

 {\bf The case $\gamma^2=1$: hyperbolic electrodynamics}
 
 Along the lines followed in Section \ref{polar} for this case, we consider first the expression (\ref{polarform}), with $\eta=constant$; the vanishing 
of the interaction terms of the form $B^{\mu}\cdot\partial_{\mu} (\xi,\eta,\theta ,\chi)$ in the Eq. (\ref{local1}) requires the identification $\theta(x) = \xi(x)$, and the fixing of the hyperbolic parameter $\chi =constant$;
\begin{eqnarray}
     {\cal L} (\psi +\psi_{0}, A+A_{0})\!\! & = & \!\! - \frac{1}{4} F^{2} -ij\frac{6e^{2}m^2_{H}}{a\lambda}B^{2} + 2\gamma ij[(\partial \rho)^2+aijm^2_{H}\rho^2]+ {\rm higher \ terms};
     \label{hypqed}
\end{eqnarray}
 Hence, the field $\xi(x)$ has disappeared and represents a Nambu-Goldstone boson in the two-dimensional valley. Therefore, as opposed to the global SSB scenario described in Eq. (\ref{rhomass}), the field $\xi$ has no survived.
Furthermore, in an orthogonal valley direction, we have a massive mode $\rho$, which has survived the gauging of the global SSB to the local symmetry. 

Similarly, in the case with a field of the form  (\ref{polarform}), with $\xi=constant$, the quadratic terms in the Lagrangian has essentially the same form shown in Eq. (\ref{hypqed}), where the interaction terms, and the terms involving the fields  $(\chi(x) , \eta(x))$ vanish by fixing now $\theta =constant$, and identifying $\chi(x) = \eta(x)$. 

Now, using the parametrization (\ref{valley}), which encodes all fluctuations in the phases, the identifications $\theta(x)  = \xi(x)$, and $\chi(x) =\eta(x)$
 lead to the vanishing of the interaction terms  in Eq. (\ref{local1}), and all terms involving the scalar fields;  the Lagrangian  reduces  to second order to a  massive pure electrodynamics,
\begin{equation}
     L = -\frac{1}{4} F^{2} -ij\frac{6e^{2}m^2_{H}}{a\lambda}B^{2} + {\rm higher \ terms};
     \label{purely}
\end{equation}
thus, the gauge vector boson has eaten the two Nambu-Goldstone bosons $(\xi ,\eta)$ and acquired a mass; the field excitations in the valley directions described in the Lagrangian (\ref{valley1}) have not survived the gauging of a global SSB to a local symmetry, and there are not scalar Higgs fields.

In the case of the conventional scalar electrodynamics one has the vector gauge field $A$, a complex field $\phi=\phi_1+i\phi_2$, with two real scalar fields, and the $S^1$ vacuum manifold with only one generator; thus, after the spontaneous symmetry breaking, only one scalar field can be eaten through Higgs mechanism, leaving a massive vector field and additionally a massive scalar field, as opposed to the massive pure electrodynamics at hand, with a hypercomplex field with two real fields, and a vacuum manifold with two generators.

{\bf The case $\gamma^2\neq1$: hyperbolic deformation of QED}

For this scenario of running parameters we shall consider two extremal cases; the first case corresponds to two massive modes with oscillations orthogonal to the two dimensional valley, and the second case with two redundant modes oscillating on the valley.

\begin{figure}[H]
  \begin{center}
  \includegraphics[width=.5\textwidth]{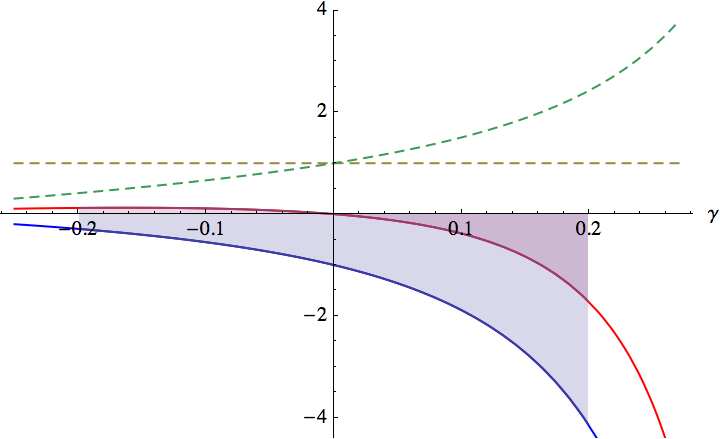}
\end{center}
\caption{The constant real mass is represented by the horizontal dashed line; the running hybrid mass is represented by the dashed curve.}
\label{twohiggs}
\end{figure}

For the first case we consider the parametrization given in Eq. (\ref{doublepolar}), and the Lagrangian reads
\begin{eqnarray}
     {\cal L} \!\! & = & \!\! - \frac{1}{4} F^{2} + e^{2}  |\psi_{0}|^{2} B^{2}+(\partial\rho_{R}+ij\partial\rho_{H})^{2} +  am^{2}(\rho_{R}+ij\rho_{H})^{2} + {\rm higher \ terms};
     \label{twomassive}
\end{eqnarray}
where
\begin{equation}
m^2 = \left \lbrace
\begin{array}{ll}
 m^2_R \Big(1+ij\frac{\gamma^2-2\gamma-1}{\gamma^2+2\gamma-1}\Big); &  (v_{0}\neq 0, w_{0}=0) \\
 m^2_R \Big(1+ij\frac{\gamma^2+2\gamma-1}{\gamma^2-2\gamma-1}\Big);  & (v_{0}= 0, w_{0}\neq 0) \end{array}\right.;
\nonumber
     \label{twomassive2}
\end{equation}
therefore, we have a Hermitian scalar field with a Hermitian mass, which has a constant real mass, and a running hybrid mass;  the first case of the above equation
the mass of the vector field $B$, is described in section (\ref{qedvacuum}), and shown in the figure (\ref{massBB}). In this case, the difference is the presence of a Hermitian Higgs field with different masses;  the figure \ref{twohiggs} shows again the masses of the field $B$ shown in the figure (\ref{massBB}), but now with the Higgs masses given in Eq. (\ref{twomassive2}). The figure shows that in this case the limit $\gamma\rightarrow \gamma_{H}$ is not a limit for light Higgs fields; however, the limit  $\gamma\rightarrow -\gamma_{H}$ corresponds again to a massive fields limit. The hybrid component of the mass is running in the interval
$m^2_{H}\in \Big(m^2_{H}(\gamma_{H}),m^2_{H}(-\gamma_{H})\Big)\approx \Big(0.4142,2.4142\Big)m^2_{R}$.
Furthermore, the second case described in the Eq. (\ref{twomassive2}) can be obtained from the first one by the transformation $\gamma\rightarrow -\gamma$,
and the corresponding figure is basically the mirrored image respect to the $``y"$-axis of the figure \ref{twohiggs}.

The case with two scalar redundant modes can be developed along the ideas behind the Eq. (\ref{purely}), leading to the Lagrangian,
\begin{equation}
     L = -\frac{1}{4} F^{2} + e^{2}  |\psi_{0}|^{2} B^{2}+ {\rm higher \ terms};
     \label{purely2}
\end{equation}
the only difference is the Hermitian mass for the vector field $B$; depending on the vacuum expectation values $(v_{0}, w_{0})$ such a running mass is described by the figure \ref{massBB}, or the mirrored figure \ref{massBB2}; in both cases there will be not dashed curves, since the scalar Higgs fields have strictly disappeared; we have again a purely massive electrodynamics, without any clues of scalar fields.

\subsection{ Local topological strings: Aharanov-Bohm-like defects}
\label{tdlocal}
The geometrical and topological description of the formation of global string defects given in section \ref{topdef} is valid in essence for the local case, adding the asymptotic form for the gauge field, and the gradients of the fields,
\begin{eqnarray}
& &  \lim_{r \to +\infty}\psi(r,z,\theta)=\rho_{0}e^{in\theta}e^{jlz}, \qquad  \lim_{r\rightarrow+\infty}D_{_{A}}\psi=0, \qquad n\in {\cal Z}-\{0\};\nonumber\\
& & \lim_{r\rightarrow+\infty} {\bf A}(r,z,\theta)=\frac{1}{e}\nabla (n\theta - ijlz)=\frac{1}{e}\Big(\frac{n}{r}\hat{\theta}-ijl\hat{z}\Big),\qquad
 \lim_{r\rightarrow+\infty} F_{\mu\nu} =0;
\label{boundary}
\end{eqnarray}
these asymptotic expressions correspond to the boundary conditions for strings solutions of nontrivial windings of a circle onto the vacuum manifold. Therefore, with these conditions,
the magnetic flux passing through a closed surface $S$ is given by
\begin{equation}
\int_{S} {\bf B\cdot dS} = \oint_{C} {\bf A\cdot dC}= -\frac{2\pi n}{e},  \quad {\bf d C} = r\hat{\theta}d\theta;
\label{quantum}
\end{equation}
where $C$ is an infinitely large loop at spatial infinity, and we have used the Stoke theorem; at this region, the vector $\hat{\theta}$ points tangentially to the loop, and $\hat{z}$ points orthogonally in the cylindrical direction. Therefore, the conventional quantized magnetic flux for string defects remains intact.

Now the ansatz includes the expression (\ref{static}) for the field $\psi$, and the following expression for the gauge field
\begin{eqnarray}
     {\bf A} (r,z,\theta) \!\! & = & \!\! \frac{i}{e} f_{A}(r) e^{in\theta}e^{jlz} \nabla (e^{-in\theta}\cdot e^{-jlz}) \nonumber \\
     \!\! & = & \!\! \frac{1}{e} f_{A}(r) \Big(\frac{n}{r} \widehat{\theta} - ijl\widehat{z}\Big), \nonumber \\
     f_{A} (0) \!\! & = & \!\! 0, \qquad f_{A} (\infty ) = 1;
     \label{anA}
\end{eqnarray}
in consistency with the asymptotic limits (\ref{boundary}). This potential results in a magnetic field
\begin{equation}
     {\bf B} = \nabla\times {\bf A} = \frac{f'_{A}}{e} \Big(\frac{n}{r} \widehat{z} + ijl \widehat{\theta}\Big),
     \label{mag}
\end{equation}
where we have a new hybrid contribution in the circular direction; note that ${\bf \overline{B}} = {\bf B}$. 

Therefore, from the expression (\ref{static}), (\ref{anA}), and (\ref{mag}), the equations of motion (\ref{eqem1}) reduce to,
\begin{equation}
     -\frac{n}{e} \frac{d}{dr} \Big(\frac{f'_{A}}{r}\Big) \widehat{\theta} + \frac{ijl}{e} \frac{1}{r} \frac{d}{dr} \Big(rf'_{A}\Big) \widehat{z} = 2e\rho^{2}_{0} f^{2} (1-f_{A}) \cdot \Big(\frac{n}{r} \widehat{\theta} -ijl\widehat{z}\Big),
\end{equation}
where the new contribution correspond to hybrid terms in the $\widehat{z}$-direction on both sides; these equations reduce explicitly to
\begin{eqnarray}
     \widehat{\theta} \!\! & : & \!\! rf''_{A} - f'_{A} + 2e^{2}\rho^{2}_{0} r(1-f_{A})f^{2} = 0, \label{flat1} \\
     \widehat{z} \!\! & : & \!\! rf''_{A} + f'_{A} + 2e^{2} \rho^{2}_{0} r(1-f_{A}) f^{2}= 0. \label{flat2}
\end{eqnarray}
Additionally, the Eq. (\ref{eqem2}) reduces to
\begin{equation}
     \frac{1}{r} \frac{d}{dr} (rf') - f(1-f_{A})^{2} (\frac{n^{2}}{r^{2}} - \underbrace{l^{2}}) + \frac{m^{2}}{2} (1-f^{2})f = 0,
     \label{eqem3}
\end{equation} 
where we have underbraced the new contribution coming from the hyperbolic phase $e^{jlz}$.
Furthermore, the Eq.(\ref{flat1}) is exactly the same equation obtained in the usual formulation for the U(1) local strings; this equation together with the corresponding equation of the form (\ref{eqem3}), have no closed-form solutions. Hence the asymptotic analysis is used for large $r$, and close to the vortex core. However, the presence of the Eq. (\ref{flat2}), distinctive of this hypercomplex formulation, has a dramatic effect on the possible solutions, enforcing the vacuum configuration for $f_{A}$ in full space, except for $r=0$;
\begin{equation}
     f_{A}(0) = 0, \qquad f_{A}(r) = 1, \qquad r \in (0, +\infty ),
     \label{abd}
\end{equation}
thus, the potential is ''pure gauge", 
\begin{eqnarray}
     \!\! & & \!\! {\bf A} (0)=0; \qquad {\bf B} (0) = 0, \quad r=0, \quad {\rm in \ the \ vortex \ core}, \label{vc} \\
     \!\! & & \!\! {\bf A} = \frac{1}{e} \Big(\frac{n}{r} \widehat{\theta} - ijl\widehat{z}\Big); \qquad {\bf B} (r) =0, \qquad r\in (0, +\infty ), \quad {\rm elsewhere}; \label{outsidevc}
\end{eqnarray}
this closed-form solution shows two representative features of a topological defect, namely the field configurations where the symmetry is left unbroken, and the configurations at the vacuum, where the symmetry will be spontaneously broken; note from the Eq.  (\ref{outsidevc}) that the potential ${\bf A}$ can not vanish at the vortex core, due to the restriction $n\neq 0$.
The presence of ``pure gauge" potentials in spaces with nontrivial topology, invokes immediately the Aharanov-Bohm effect \cite{ab}: a narrow, infinite length solenoid is added to the two-slit experiment for electrons; topologically the solenoid is a string-like defect, and thus the phase shift on the electron wave function is a topological effect determined by the magnetic flux inside the solenoid. Outside, there is no magnetic field, with non-zero potential. Since the original experimental confirmation \cite{chambers}, the Aharanov-Bohm effect has received considerable study, from theoretical generalizations, to varied experimental realizations; for example, the appearance of the electronic interference phenomenon in carbon nanotubes suggests that the Aharanov-Bohm effect is relevant even at the microscopic scale \cite{abexp}. With this perspective, we establish now the analogy with the case at hand.

In the analogy, the infinite string core corresponds to the Aharanov-Bohm solenoide; the a\-zi\-mu\-thal component of ${\bf A}$ in the expression (\ref{outsidevc}) falls off like $1/r$, with the distance from the core, such as in the Aharanov-Bohm effect. Additionally the new contribution in the $\hat{z}$-direction is constant everywhere; this new contribution does not change the quantization condition of the magnetic flux in (\ref{quantum}). Now an important difference; in the core of the string defect the magnetic field vanishes, as opposed to the Aharanov-Bohm solenoide, inside which the ${\bf B }$ field is non-zero; the string defect at hand is actually a `pure gauge" phenomenon. Furthermore, in the background, one has the same topological feature, namely, the existence of a non-simply connected space, and thus the winding number around the loop is observable in the Aharanov-Bohm effect; in the analogy, this fact  represents a phenomenological possibility for the string defects at hand.

Let us see how these ``pure gauge" potentials are able to make finite the energy of the defect, playing the role of compensating fields for the divergent effect of the scalar fields.

The substitution of the solution (\ref{abd}) into the Eq.(\ref{eqem3}) leads to a simplified equation,
\begin{equation}
     \frac{1}{r} \frac{d}{dr} (rf') + \frac{m^{2}}{2} (1-f^{2}) f = 0;
     \label{simplified}
\end{equation}
an immediate solution is $f=1$, and thus, the field $\psi$ settles down to its vacuum configuration in the full space, except in the core. Note that this solution for $f$ does not solve the Eq. (\ref{f}) for global strings. In this case, all components of the energy-momentum tensor (\ref{emtqed}) vanish trivially, and hence are finite.
Therefore, all classical Maxwell physical observables vanish, and there is no classical experiment that allows to detect the string defect by its electrodynamical effects. However, the loop integral (\ref{quantum}) is a gauge invariant quantity, and according to the Aharanov-Bohm effect, is detectable for quantum interference.

\subsection{ There are not other solutions for local strings}
\label{otherstrings}
 The usual description of local string defects that involves only the Eqs. (\ref{flat1}) and (\ref{eqem3}), the asymptotic solutions are well known \cite{vilenkin}; hence, close to the vortex core, one has that,
\begin{equation}
     f_{A} \sim r^{2}, \qquad f\sim r^{n}, \qquad r\rightarrow 0;
     \label{vilen1}
\end{equation}
and for large $r$,
\begin{equation}
     1-f_{A} \sim e^{-mr}, \qquad 1-f \sim e^{-\beta r}, \qquad \beta = \frac{\lambda}{e^{2}}, \qquad r\rightarrow\infty ;
     \label{vilen2}
\end{equation}
physically, these expressions lead to an energy density more localized that in the global strings, and show that the asymptotic behaviour of $f$ is controlled by the gauge field contribution $f_{A}$. These expressions are not valid in the present treatment, since $f_{A}$, has been relaxed to its vacuum configuration in the full intervale $r\in (0,+\infty)$ due to the Eq. (\ref{flat2}); we have only at hand the Eq.(\ref{simplified}), which is fully independent on $f_{A}$.

Other closed-form solutions for Eqs. (\ref{simplified}) are not known; however, one can obtain asymptotic solutions for large $r$ and close to the core using linearized versions; for large $r$ the background field is $f=1$, and close to the vortex $f=0$;
\begin{equation}
     r\triangle f'' + \triangle f' - m^{2}r\triangle f=0, \qquad r\rightarrow\infty, \qquad \triangle f= AJ_{0} (imr) + BN_{0} (-imr),
     \label{abasymp}
\end{equation}
and
\begin{equation}
     r\triangle f'' + \triangle f' + \frac{m^{2}}{2} r\triangle f=0, \qquad r\rightarrow 0, \qquad \triangle f = A J_{0} \Big(\frac{m}{\sqrt{2}}r\Big) + BN_{0} \Big(\frac{m}{\sqrt{2}}r\Big),
     \label{abasymp2}
\end{equation}
where $J_0$ is the zero-order Bessel function, $N_0$ the zero-order Neumann function, and
 $A$ and $B$ are adimensional constants. However, in the Eq.\ (\ref{abasymp2}), the condition $A=0$ is enforced for ensuring a single-valued function at $r=0$; thus, the energy density is given essentially by the expression
\begin{equation}
T_{00}= -\frac{1}{2}(\triangle f')^2= -\frac{B^2 m^2}{4}\Big[N_1\Big(\frac{m}{\sqrt{2}}r\Big)\Big]^2,
\label{neu1}
\end{equation}
where $N_{1}$ is the one-order Neumann function; thus, the energy diverges, similarly to the case of global strings. Since local strings have their energy confined mainly close to the core, such a solution is not physically meaningful; in this case the gauge field $f_{A}$ has desapeared as a compensating field of the divergent effects of the scalar field. Therefore, the Aharonov-Bohm like defects discussed early, are the only field configurations with finite energy.
\section{Concluding remarks} 
\label{cr}
 \subsection{Cosmological implications}
 \label{cosmos}
The results obtained can be interpreted in various senses; in the inflationary cosmology context, the topological defects are essential in the spontaneous symmetry breaking based phase transitions; if a phase transition occurs, then they may be generated provided that the vacuum manifold has a nontrivial topology. Since a dramatic effect of the substitution of the circular rotations by the hyperbolic rotations is the trivialization of the vacuum manifold from the homotopic point of view (section \ref{gamma1}), then the insignificant observational support on the existence of cosmic string defects leads to the possibility that the vacuum can manifest a non-compact topology in certain phase of the early-universe. This in its turn has various implications; for example it suggests that the GUT phylosophy must be extended by incorporing non-compact gauge groups.
Furthermore, the proliferation of cosmic defects is a essential feature of certain GUT's; {\it inflation} was originally proposed as a form of explaining the observational evidence against such a proliferation. At the light of the present results, one can alternatively to postulate the hyperbolic symmetry as a essential symmetry at some period of the early-universe; thus, the inflationary scenarios can be modified drastically by the presence of the new symmetry. More explorations are mandatory 
along these ideas.
 
\subsection{On Aharonov-Bohm strings}
\label{bohm}

It has been shown that the dominant interaction between matter and cosmic strings is through an Aharonov-Bohm interaction \cite{alford0}; outside the tiny region of the inner core, the field strengths vanish, but one has non-vanishing potentials in the outer region,  and hence a scattering mechanism of the Aharonov-Bohm type will work at the outer region. The scattering cross sections and production rates do no go to zero as the geometrical size of the string goes to zero. One can realize that the results obtained in the present approach are consistent with those results, but suggesting rather that the Aharonov-Bohm interaction is the only interaction between matter and cosmic strings; the field configurations for the strings found in the section (\ref{tdlocal}) do not distinguish between inner core and outer region, since the potentials are pure gauge as close to the core as one wants.

The Aharonov-Bohm strings have appeared previously in the study of discrete gauge symmetries  \cite{krauss, alford, preskill,seiberg}; specifically in the effective Lagrangian description of $Z_{k}$ discrete gauge theory, this type of strings have confined magnetic flux with a $1/k$ unit of fundamental magnetic charge;  due to this charge the Aharonov-Bohm string can interact with matter. Recently,  certain cosmological constraints  have been imposed on these theories, by studying the radiation of standard model particles from these strings \cite{yutaka}. As opposed to these discrete gauge symmetry approaches, the Aharonov-Bohm-like strings have been obtained in the approach at hand by incorporating a (non-compact) continuous symmetry, the hyperbolic rotations; it may be interesting to study  phenomenological models that incorporate both, discrete, and noncompact gauge symmetries, following the ideas des\-cri\-bed in \cite{yutaka}.

{\bf Acknowledgements:}
This work was supported by the Sistema Nacional de Investigadores (M\'exico), and the Vicerrectoria de Investigaci\'on y Estudios de Posgrado
(BUAP). The authors would like to thank Dra. I. Rubalcava-Garcia for discussions. The numerical analysis and graphics have been made using Mathematica.

\end{document}